\documentclass[aps,onecolumn,groupedaddress,amsmath,amssymb,nofootinbib]{revtex4}
\usepackage{amsmath, dsfont}
\usepackage{amssymb}
\usepackage{epsfig}
\usepackage{calc}
\usepackage{amsfonts}
\usepackage{graphicx}
\usepackage[ansinew]{inputenc}
\usepackage{multirow}
\usepackage{subcaption}
\usepackage{tensor}

\usepackage{natbib}

\begin{document}

\title[]{Selecting Horndeski theories without apparent symmetries and their black hole solutions}

\author{Eugeny Babichev$^{\dagger}$, Christos Charmousis$^{\dagger}$, Mokhtar Hassaine$^{\ddag}$ and Nicolas Lecoeur$^{\dagger}$}
\affiliation{$^{\dagger}$Universit\'e Paris-Saclay, CNRS/IN2P3,
IJCLab, 91405
Orsay, France,\\
$^{\ddag}$Instituto de Matem\'atica, Universidad de Talca, Casilla
747, Talca, Chile.}

\begin{abstract}
Starting from a generalised Kaluza-Klein action including arbitrary Horndeski potentials, we establish integrability and compatibility conditions that solve the generic field equations for spherical symmetry. The resulting theories can be identified as general Horndeski theories having no apparent symmetries in four dimensions or as effective string theory actions with an IR logarithmic running for the dilaton, higher order corrections and generalised Liouville type potentials. For such actions, we then find black holes with secondary hair parameterized by two coupling constants essentially characterising the theories at hand. One is related to an action which is conformally coupled in five dimensions while the second is related to a Kaluza-Klein reduction of Lovelock theory. We show that the full action can also be interpreted as a sum of conformally coupled actions in differing dimensions. Known solutions are mapped within the general chart of the found theories and novel general black holes are discussed, focusing on their important properties and some of their observational constraints.
\end{abstract}

\maketitle

\section{Introduction}
General Relativity (GR) is quite successful in passing numerous
strong field tests that recent experimental and observational
advances have brought forward. Amongst them we can mention the
direct detection of gravitational waves emitted by binary inspirals
(see e.g.~\cite{LIGOScientific:2017vwq}), spectroscopy of X-rays
produced by accreting black holes (see e.g.~\cite{Remillard:2006fc}), observation of star(s) trajectories
orbiting the supermassive black hole Sagittarius $A^*$ at the center
of the Milky Way \cite{GRAVITY:2018ofz} and the recent direct
observation of black holes by the Event Horizon Telescope
\cite{EventHorizonTelescope:2020qrl}. These international
collaborative efforts consolidate GR in the strong gravity regime.
It goes without saying that recent observations also raise
questions. A typical example is the  mass of the secondary in the
event GW190814, which remains puzzling in the framework of GR. Its
mass of $2.59^{+0.08}_{-0.09}$ solar masses is too light for a black
hole, and too heavy for a neutron star with usual equation of state
or angular momentum (and in accordance with the multi messenger
observation of \cite{LIGOScientific:2017vwq} disfavoring stiff
equations of state). This is even more intriguing as it is for small
masses that we may hope or expect UV effects beyond GR since the
horizon or surface of the compact object are strongly curved. With
expected upcoming observations of increasing precision, it is
paramount to find compact objects of modified gravity theories,
which satisfy the previously mentioned tight experimental
constraints and can also point beyond GR. Scalar tensor theories can for
example modify the no-hair relation of GR and admit black hole
solutions which differ from the Kerr spacetime (see for example
\cite{Anson:2020trg}), providing a theoretical framework to quantify
possible observational deviations from GR. Furthermore, within scalar-tensor theories it was for example found in~\cite{Charmousis:2021npl} that there exist neutron star solutions
with no mass gap permitting more massive neutron stars (of generic
equation of state) or lighter black hole secondaries as for example
detected in GW190814. These results, amongst others, are not as yet
conclusive but interesting in their own right as a measurable
departure from GR.

It is not surprising that most analytic results concerning compact
objects in such theories were obtained making some assumption(s) of
symmetry for the scalar-tensor action at hand. Early on
stealth/self-tuning solutions of spherical symmetry were obtained
\cite{Babichev:2013cya} under parity $\phi\to -\phi$ and shift
symmetry $\phi\to\phi+\text{cst}$. They were extended for generic
classes of scalar-tensor theories~\cite{Kobayashi:2014eva}. These
results were even extended to stealth stationary metrics
\cite{Charmousis:2019vnf}, upon realising the intricate relation of
the scalar with the Hamilton-Jacobi functional of the Kerr metric
described in the classic works of B. Carter~\cite{Carter:1968ks}.
For theories with broken parity or/and broken shift symmetry,
numerical studies were initially made demonstrating non-stealth
configurations (see for example
\cite{Sotiriou:2014pfa,Babichev:2016fbg}), non-perturbative effects
like scalarisation~\cite{Doneva:2017bvd} or scalar shells around
black holes~\cite{Babichev:2022djd}. On the analytic side, using a generalised Kaluza-Klein dimensional
reduction~\cite{Kanitscheider:2009as} of Lovelock theory, one could
construct black hole solutions \cite{Charmousis:2012dw} with higher
order corrections and without shift or parity symmetry (see also more recently \cite{Bakopoulos:2022csr}). Such a
Kaluza-Klein truncation involved all Horndeski scalars for given
exponential type potentials. These solutions were, from a string
theory point of view,  the leading $\alpha'-$correction of two
derivative Einstein dilaton solutions with a Liouville potential
(initially discussed in \cite{Chan:1995fr}). At the absence of mass,
a naked singularity was present for the two derivative Einstein
dilaton theories. When these were completed to higher order theories
however, the naked singularity was cloaked by an event horizon
originating from the higher order terms of the action
\cite{Charmousis:2012dw}. However, the mass term of such
$\alpha'-$corrected solutions did not have a Newtonian fall off.
This was in agreement with their higher dimensional Lovelock theory
origin. This drawback was remedied very recently. Indeed, analytic
solutions can be found for a precise combination of all Horndeski
scalars \cite{Fernandes:2021dsb} upon taking an intriguing four
dimensional singular limit (initially considered in
\cite{Glavan:2019inb}). For the theory in question
\cite{Fernandes:2021dsb}, symmetries are still present but now
somewhat hidden, the scalar field equation of motion being
conformally invariant, while the scalar-tensor action is not (see also~\cite{Lu:2020iav,Babichev:2022awg} and the review \cite{Fernandes:2022zrq}). The
Gauss-Bonnet curvature scalar, a crucial term of Lovelock theory and
in this construction, is given by
$\mathcal{G}=R^2-4R_{\mu\nu}R^{\mu\nu}+R_{\mu\nu\rho\sigma}R^{\mu\nu\rho\sigma}$.
In four dimensions, it is a topological invariant (the generalised
Euler density), unless coupled with a scalar (see for example the
nice discussion in \cite{Sotiriou:2014pfa}). The above mentioned
theories \cite{Fernandes:2021dsb,Charmousis:2012dw} are higher
dimensional Kaluza-Klein truncations of Lovelock gravity. They can
also be interpreted as low-energy effective holographic theories \cite{Charmousis:2010zz} at
large but finite couplings or non critical string theory at finite
temperature backgrounds (see eg., \cite{Dixon:1986iz}) with leading
string tension $\alpha'$-UV corrections. The common denominator is
the presence of the scalar or dilaton field driving the low energy
dynamics in the IR. In a nutshell generalised stringy Kaluza-Klein
compactifications lead to the appearance of generic Horndeski  terms
in four-dimensional actions with particular couplings for the scalar
field. So, can one go any further in this promising direction?

All things considered, there is no straightforward physical reason
to have some symmetric configuration for the scalar-tensor theory.
Our aim in this paper will be to strive a step further and obtain
black hole solutions for scalar-tensor theories without some obvious
symmetry present, including a generic form of Horndeski theory with
no shift, parity nor partial conformal symmetry (for a recent brief review see \cite{Bakopoulos:2022gdv}). A good starting
point towards this aim is to try to generalise the potentials
appearing in Kaluza-Klein compactifications originating from
Lovelock theory \cite{Charmousis:2012dw}. Such actions include all
Horndeski terms with the scalar couplings depending on the nature of
the compactified extra dimensional space. Here we should pause in
order to emphasize a key difference of Lovelock theory to higher
dimensional GR first noted in \cite{Dotti:2005rc}. Whereas locally
compact spaces, present as horizons, are Einstein spaces in GR, for
Lovelock theory this property is not sufficient. Horizons must in
addition have constant Gauss-Bonnet curvature, a rather strong
condition as it turns out. This reduces the possible spaces upon
which to compactify Lovelock black hole solutions. In other words,
uniform black strings or black branes are not solutions in Lovelock
theories. A notable non trivial example remains; the product of an
arbitrary number of two-spheres giving rise to a black hole solution
in higher dimensions \cite{Bogdanos:2009pc}. It is this solution
that truncates to the Kaluza-Klein solution found early on in
\cite{Charmousis:2012dw}. It is the  singular limit of this solution
to four dimensions which gives one of the black hole solutions found
independently by Fernandes \cite{Fernandes:2021dsb}.

Given these considerations, let us consider the following action
which includes Kaluza-Klein potentials and particular subcases with
known solutions, see e.g.
\cite{Charmousis:2012dw,Fernandes:2021dsb}:
\begin{equation}
\begin{split}
S = \int \mathrm{d}^4x\sqrt{-g}\left\lbrace \left(1+W\left(\phi\right)\right)R-
\frac{1}{2}V_k\left(\phi\right)\left(\nabla\phi\right)^2+Z\left(\phi\right)+
V\left(\phi\right)\mathcal{G}+V_2\left(\phi\right)G^{\mu\nu}\nabla_\mu\phi\nabla_\nu\phi\right.\\
\left.  +V_3\left(\phi\right)\left(\nabla\phi\right)^4+V_4\left(\phi\right)\Box\phi\left(\nabla\phi\right)^2\right\rbrace.
\end{split} \label{eq:action}
\end{equation}
Note the presence of a canonical kinetic term for the scalar field
which is important in order to evade possible strong coupling
issues. Then note the potential $W(\phi)$ which determines if the
scalar is minimally coupled to the Ricci scalar (to lowest order) or
not. The potential $V\left(\phi\right)$ multiplying the Gauss-Bonnet
invariant is defined up to an additive constant which would only
yield a boundary term. The remaining higher order potentials are
subsets of $G_4, G_2$ and $G_3$ potentials of Horndeski theory
respectively. Finally, the potential term $Z(\phi)$ includes the
cosmological constant, Liouville exponential terms most commonly
present in non critical string theories \cite{Dixon:1986iz}, in self tuning scenarios \cite{Charmousis:2017rof},
and holographic gravitational backgrounds (see
\cite{Charmousis:2012dw} and references within).
The theory as a whole therefore includes string effective theories
with higher order $\alpha'-$corrections. It also includes the
typical Kaluza-Klein compactification one would get from Einstein
and Gauss-Bonnet higher dimensional gravity \cite{Charmousis:2012dw}. To these
considerations, we finally add the interesting action considered by
Fernandes \cite{Fernandes:2021dsb}, which fits in the framework of
action (\ref{eq:action}) with the following potentials,
\begin{equation}
W = -\beta\mathrm{e}^{2\phi},\quad V_k = 12\beta\mathrm{e}^{2\phi},\quad Z = -2\lambda \mathrm{e}^{4\phi}-2\Lambda,
\quad V = -\alpha\phi,\quad V_2 = 4\alpha = V_4 ,\quad V_3 = 2\alpha,\label{eq:pot_fern}
\end{equation}
with three coupling constants $\alpha$, $\beta$ and $\lambda$. In
his case the scalar field equation is conformally invariant,  and as
such led to an interesting  universal geometric constraint. This is
no longer true for the general action under present consideration.

We are interested in static and spherically symmetric spacetimes. The following metric ansatz is considered throughout this paper for spherical symmetry,
\begin{equation}
\label{metric}
\mathrm{d}s^2 = -f\left(r\right)\mathrm{d}t^2+\frac{\mathrm{d}r^2}{f\left(r\right)}+r^2\mathrm{d}\Omega^2,\quad \phi=\phi\left(r\right).
\end{equation}
This is not the general spherically symmetric ansatz. Therefore we
will have to be cautious on the compatibility of the field equations
for the theory at hand. We will now seek a way to filter out
theories reducing the above general action to a more tractable yet
quite general theory to solve. The most interesting results of our
analysis are obtained when $\phi(r)$ is not constant, which we
assume from now on, while we treat the case of constant scalar field
in appendix~A.

The paper is organized as follows: in Sec. II, we present conditions
on the potentials which enable to rewrite the field equations as
three simple compatibility conditions, parameterized by a unique
real function $\mu(r)$. Sec. III contains the main results, namely
two novel black hole solutions obtained for the case $\mu(r)\equiv
1$ and with Newtonian mass fall off. In Sec. IV, the more general
case of $\mu(r)\equiv\mu=\text{cst.}$ is treated, also leading to
static black hole solutions, while Sec. V is devoted to our
conclusions.

\section{Integrability and compatibility}\label{sec:int_comp}

In this section, we will see that a specific truncation of
(\ref{eq:action}) can accommodate analytic solutions for the ansatz
(\ref{metric}). For this purpose, let us denote by
$\mathcal{E}_{\mu\nu} = \frac{\delta S}{\delta g^{\mu\nu}}$ the
field equations arising from the variation of the action with
respect to the metric. Given a homogenous ansatz, (\ref{metric}), it
is quite common to consider the combination
$\mathcal{E}^t_t-\mathcal{E}^r_r=0$ in order to determine the
expression of the scalar field. We obtain,
\begin{equation}
\begin{split}
\frac{\phi''}{\left(\phi'\right)^2}\left[r^2W_\phi +
4\left(1-f\right) V_\phi + 2frV_2 \phi' + fr^2
V_4\left(\phi'\right)^2\right] =
-\frac{r^2}{2}\left(V_k+2W_{\phi\phi}\right) -
\left(V_2+4V_{\phi\phi}\right)\left(1-f\right)\\ -
fr\left(V_{2\phi}-2V_4\right)\phi' - fr^2
\left(V_{4\phi}-2V_3\right)\left(\phi'\right)^2, \label{eqwfactor}
\end{split}
\end{equation}
where prime stands for radial derivative, while a subscript
$\phi$ denotes a derivation with respect to $\phi$. Upon close inspection, we see that
choosing the potentials of the theory as follows,
\begin{equation}
V_k+2W_{\phi\phi} = \frac{2}{d(\phi)}W_\phi,\quad V_2+4V_{\phi\phi}
= \frac{4}{d(\phi)}V_\phi,\quad V_{2\phi}-2V_4 =
\frac{2}{d(\phi)}V_2,\quad V_{4\phi}-2V_3 = \frac{1}{d(\phi)}V_4,
\label{conditions}
\end{equation}
equation (\ref{eqwfactor}) is factorized in a simple and elegant
way providing {\it a priori} two branches of
solutions,
\begin{equation}
\Biggl[\frac{\phi''}{\left(\phi'\right)^2}+\frac{1}{d(\phi)}\Biggr]\Biggl[r^2W_\phi
+ 4\left(1-f\right) V_\phi + 2frV_2 \phi' + fr^2
V_4\left(\phi'\right)^2\Biggr]=0. \label{eq:simple_equation}
\end{equation}
It is important that the conditions (\ref{conditions}) only involve
the potentials of the theory---they cannot involve the
characteristics of our ansatz (\ref{metric}) or the seeked-for
solutions. Here $d=d(\phi)$ is some non-vanishing function of the
scalar, and we will see below that, for the first branch of
solutions (\ref{eq:simple_equation}), it can be chosen  to be
constant without any loss of generality. It is interesting to note
that, under the conditions (\ref{conditions}), the potentials $V_k$
and $V_i$ for $i=2, 3$ and $4$ can be parameterized in terms of the
Einstein-Hilbert and Gauss-Bonnet potentials $W$ and $V$ as
\begin{align}
V_k = {}& \frac{2}{d}W_\phi-2W_{\phi\phi},\label{eq:vk}\\
V_2 = {}& \frac{4}{d}V_\phi-4V_{\phi\phi},\\
V_4 ={}& -\frac{2}{d^2}\left(2+d_\phi\right)V_\phi+\frac{6}{d}V_{\phi\phi}-2V_{\phi\phi\phi},\\
V_3 ={}&
\frac{1}{d^2}\left[\frac{1}{d}\left(1+2d_\phi\right)\left(2+d_\phi\right)-d_{\phi\phi}\right]
V_\phi-\frac{1}{d^2}\left(5+4d_\phi\right)V_{\phi\phi}+\frac{4}{d}V_{\phi\phi\phi}-V_{\phi\phi\phi\phi}.
\label{eq:v3}
\end{align}
In other words, the factorization (\ref{eq:simple_equation}) is made
possible with any action (\ref{eq:action}) parameterized by three
independent potentials, namely $W$, $Z$ and $V$, provided that the
remaining potentials are fixed by the above equations.

Now, let us take a closer look at the factorization
(\ref{eq:simple_equation}) and at its possible consequences for our
purpose. First of all, the potentials
(\ref{eq:pot_fern}) of Ref. \cite{Fernandes:2021dsb} correspond to a
constant function $d$ given by $d\left(\phi\right)=-1$. As it also
occurs in this latter reference, the equation
(\ref{eq:simple_equation}) offers the possibility of two branches of
solutions for the scalar field. We note that the first branch
corresponding to the first bracket in (\ref{eq:simple_equation})
does not involve coupling functions of the theory nor the
metric function. We have a simple ODE giving the scalar field independently of the geometry{\footnote{This is typical in stringy black hole solutions with an IR logarithmic running for the dilaton (see for example \cite{Chan:1995fr}, \cite{Charmousis:2010zz}).}}.
The second branch is much more involved because
the equation involves explicitly the coupling potentials of the
theory and the metric function. We discuss this latter case in Appendix B. We here focus on the first branch for which the scalar
field satisfies the equation
\begin{equation}
\phi'' =
-\frac{\left(\phi'\right)^2}{d\left(\phi\right)}.\label{eq:first_branch}
\end{equation}
To go further we now need to show the compatibility of the remaining
equations with (\ref{eq:first_branch}) and our ansatz for the
metric~(\ref{metric}). This requires fixing the potentials $W$, $Z$
and $V$ in such a way that the two remaining independent equations
admit the same metric function $f$ as solution. Again, the
potentials, which are functions of $\phi$ or equivalently of $r$
via~(\ref{eq:first_branch}), must be independent of the metric
solution $f$. It is quite remarkable that taking into account the
expression of $\phi''$ from (\ref{eq:first_branch}), the two
independent equations, $\mathcal{E}_{rr}=0$ and
$\mathcal{E}_{\theta\theta}=0$ can be integrated once and twice
respectively to give
\begin{equation}
\mathcal{E}_{rr} \propto I_1'\left(r\right),\quad
\mathcal{E}_{\theta\theta}\propto I_2''\left(r\right),
\label{eq:errethth}
\end{equation}
with
\begin{align}
I_1\left(r\right)={}& f^2\left(r^2V\right)'''-f\left(2r\left(1+\mathcal{W}'\right)+4V'+r^2\mathcal{W}''\right)+2r+2\mathcal{W}+r\mathcal{Z}'-\mathcal{Z},\\
I_2\left(r\right)={}& f^2\left(r V\right)'' - f
r\left(1+\mathcal{W}'\right) + \mathcal{Z},
\end{align}
and, where we have introduced for clarity two auxiliary functions $\mathcal{W}$ and
$\mathcal{Z}$ determined by,
\begin{equation}
W = \mathcal{W}',\quad rZ = \mathcal{Z}''. \label{eq:wz}
\end{equation}
The integration of the equations (\ref{eq:errethth}) implies the
existence of three integration constants, $d_1$, $c_1$ and $c_2$, such that
\begin{equation}
I_1 - d_1 = 0,\qquad I_2 -c_2+c_1 r =0. \label{eq:eqforf}
\end{equation}
As the following calculations show, the integration constants $c_1$, $c_2$, $d_1$ are
not independent and are either gauged away or related to the mass of the
black hole. Compatibility of the field equations is ensured, once the two quadratic
equations $I_1$ and $I_2$, defining the metric function $f$, are
proportional. Denoting by $2\mu\left(r\right)$ this proportionality
factor, which is \textit{a priori} an arbitrary non-vanishing function of $r$, one obtains the following system of
equations,
\begin{align}
\left(r^2 V\right)'''  ={}& 2\mu \left(r V\right)'',\label{eq:comp1}\\
4 V' = {}& 2 \left(\mu-1\right)r\left(\mathcal{W}'+1\right)-r^2\mathcal{W}'',\label{eq:comp2}\\
2r+2\mathcal{W} ={}& d_1-2\mu c_2+2\mu
c_1r+\left(2\mu+1\right)\mathcal{Z}-r\mathcal{Z}',\label{eq:comp3}
\end{align}
where we assumed that the factors in front of different powers of
$f$ should be proportional independently. For a given
proportionality factor $\mu\left(r\right)$, these equations will
determine the unfixed potentials $W$, $Z$ and $V$ as functions of
$r$ or equivalently of $\phi$ while the quadratic equations
(\ref{eq:eqforf}) will give the metric function $f(r)$. Note that
the parameters of the solution, in our case the mass, must not
affect the relations of the theory potentials as then the theory
would be fine tuned. We will see that this is not the case here.

As one may notice, the above conditions for compatibility of the
equations are independent of the choice of  $d(\phi)$, indicating
that changing $d(\phi)$ does not change the physical results.
Indeed,  for any scalar field satisfying~(\ref{eq:first_branch}),
the redefined scalar $\phi\to \int H(\phi)\phi' dr$
satisfies~(\ref{eq:first_branch}) with $d(\phi)=-1$, provided that
$H$ solves the ordinary differential equation
$H_{\phi}-H^2-\frac{H}{d}=0$. One can therefore, without any loss of
generality, fix $d\left(\phi\right)=-1$. Then, the general solution
of Eq.~(\ref{eq:first_branch}) is,
\begin{equation}
\phi\left(r\right) = \ln \left(\frac{c}{r+\tilde{c}}\right), \label{eq:ccprime}
\end{equation}
where $c$ and $\tilde{c}$ are two integration constants. Note that the
constant $\tilde{c}$ can be further fixed to have a specific value for
convenience\footnote{Clearly, different choice of $\tilde{c}$
in~(\ref{eq:ccprime}) amounts to a redefinition of the scalar field.
We can use the residual freedom to redefine the scalar by choosing
the constant of integration $\tilde{c}$. Indeed the function $H(\phi)$ that
provides $d(\phi)=-1$ is defined up to an integration constant,
since it satisfies $H_{\phi}-H^2-\frac{H}{d}=0$. One can show that
by adjusting this integration constant one can change $\tilde{c}$.}. In
particular, one can choose $\tilde{c}\propto c$, as in the examples just
below, or $\tilde{c}=0$ for the solutions presented in the following
sections.

Let us first demonstrate how one can reproduce certain known
solutions by using our formalism. If the Gauss-Bonnet potential
$V=0$, then (\ref{eq:comp1}) is satisfied automatically, and
Eqs.~(\ref{eq:vk})--(\ref{eq:v3}) show that the action only has the
Einstein-Hilbert potential $W$, the kinetic potential $V_k$ and the
self-interaction and cosmological constant in $Z$. In this case
therefore all higher order terms in the action (\ref{eq:action}) are
missing and we are left with an action with at most two derivatives.
This  encompasses the Bocharova-Bronnikov-Melnikov-Bekenstein (BBMB)
\cite{BBM} and the Martinez-Troncoso-Zanelli
(MTZ)~\cite{Martinez:2002ru} black holes. Indeed,
consider the following potentials,
\begin{equation}
V=0,\quad W=-\beta\mathrm{e}^{2\phi},\quad Z = -2\lambda\mathrm{e}^{4\phi}-2\Lambda,\quad V_k = 12\beta\mathrm{e}^{2\phi},
\label{chile}
\end{equation}
where $V_k$ is determined by $W$ according to (\ref{eq:vk}). Then,
we take into account the scalar field profile~(\ref{eq:ccprime}),
and solve the compatibility
conditions~(\ref{eq:comp2})--(\ref{eq:comp3}). Finally finding the
metric function $f$ from~(\ref{eq:eqforf}) we get the solution
\begin{equation}
\phi = \ln\left(\frac{M}{\sqrt{\beta}\left(r-M\right)}\right),\quad f = \left(1-\frac{M}{r}\right)^2-\frac{\Lambda r^2}{3}, \label{eq:bbmbmtz}
\end{equation}
provided the relation $\lambda=-\Lambda\beta^2$ holds.
Eq.~(\ref{eq:bbmbmtz}) gives either the BBMB black hole (for
$\Lambda=0$) or the MTZ black hole (for $\Lambda\neq 0$), with a
unique integration constant $M$ playing the role of the black hole
mass. The value of the function $\mu(r)$ in both these cases is
$\mu(r)=1+M^2/\left(2M^2-3Mr+r^2\right)$.

A known solution for non-zero $V$ is given
in~\cite{Fernandes:2021dsb}, for the theory with the
potentials~(\ref{eq:pot_fern}) with $\lambda=\beta^2/(4\alpha)$. In
our formalism this solution corresponds to the choice
$c=\sqrt{-2\alpha/\beta}$ and $\tilde{c}=0$ in the solution for $\phi$ in
Eq.~(\ref{eq:ccprime}), and the constant value of
$\mu(r)=1$\footnote{It is interesting to note that in this case,
the couplings $c$, $\tilde{c}$ and $\mu(r)$ are independent of the mass
integration constant $M$. This is unlike the BBMB and MTZ cases
where it is important that the action functions do not end up
depending on the parameter of the solution $M$, as is indeed the
case, (\ref{chile}).}.

As we mentioned above, we have a choice to fix the constant of
integration $\tilde{c}$ in~(\ref{eq:ccprime}). For the BBMB and MTZ
solutions we took $\tilde{c}\neq 0$  in order to stick with the standard
form of these solutions, while to demonstrate the solution
of~\cite{Fernandes:2021dsb} we chose $\tilde{c}=0$, again to be in accord
with the presentation of the original paper. From now on, we will
set $\tilde{c}=0$ and thus we consider
\begin{equation}
\phi\left(r\right) = \ln \left(\frac{c}{r}\right),
\label{eq:choice_of_phi}
\end{equation}
where $c>0$ is a constant with dimension $1$. As we will see below, the
constant $c$ of the scalar field solution~(\ref{eq:choice_of_phi})
is related to the coupling constants of the theory once the compatibility conditions (\ref{eq:comp1})--(\ref{eq:comp3}) are solved.

It turns out that compatibility conditions
(\ref{eq:comp1})--(\ref{eq:comp3}) can be solved for any constant
$\mu$, the most interesting case being $\mu=1$. Indeed, while we
have seen above that the solution of Ref.~\cite{Fernandes:2021dsb} corresponds to $\mu=1$, it appears conversely that solving the
compatibility conditions with $\mu=1$ leads to a more general action
than the one considered in \cite{Fernandes:2021dsb}, acquiring new
black hole solutions with far-away Newtonian asymptotics. The
following section is dedicated to this $\mu=1$ case, while the more
general case of constant $\mu$ is explained in Sec. IV.

\section{Case of $\mu=1$: Black holes with a Newtonian fall off}

For a constant
proportionality factor $\mu\left(r\right) = \mu =
\text{cst.}$, the compatibility conditions
(\ref{eq:comp1})--(\ref{eq:comp3}) are integrable, and new explicit
solutions can be found. This will be the core of the next two sections. Different choices of constant
$\mu$ yield solutions of differing far away asymptotics, and only
for $\mu=1$ do the metric solutions have a standard four dimensional Newtonian
behaviour at infinity, i.e. $f\sim 1-2M/r-(\Lambda r^2/3)$. Here, the optional
$\Lambda$-term in parenthesis, stands for the cosmological constant if present in the action, while $M$ is the mass of the solution.

Let us thus focus in the current section on $\mu=1$, and present the results in a way which enables to interpret them easily. We consider the following potentials $W, Z$ and $V$,
\begin{equation}
W = -\beta_4\mathrm{e}^{2\phi}-\beta_5\mathrm{e}^{3\phi},\quad Z =
-2\lambda_4\mathrm{e}^{4\phi}-2\lambda_5\mathrm{e}^{5\phi}-2\Lambda,\quad
V=-\alpha_4\phi-\alpha_5\mathrm{e}^{\phi},\label{eq:pot_mu_egal_1}
\end{equation}
where $\beta_4$, $\beta_5$, $\lambda_4$, $\lambda_5$, $\alpha_4$ and
$\alpha_5$ are six coupling constants, and $\Lambda$ is the usual cosmological constant. The choice of subscripts $"4"$ and
$"5"$ will become clear momentarily. The
remaining potentials are given by the compatibility conditions
(\ref{eq:vk})--(\ref{eq:v3}) with $d=-1$, giving the following action,
\begin{align}
S = {}&\int\mathrm{d}^4x\sqrt{-g}\Biggl\{
R-2\Lambda-2\lambda_4\mathrm{e}^{4\phi}-2\lambda_5\mathrm{e}^{5\phi}-
\beta_4\mathrm{e}^{2\phi}\left(R+6\left(\nabla\phi\right)^2\right)-
\beta_5\mathrm{e}^{3\phi}\left(R+12\left(\nabla\phi\right)^2\right)
\nonumber\\{}&{}-\alpha_4\left(\phi\mathcal{G}-4G^{\mu\nu}\phi_\mu\phi_\nu-
4\Box\phi\left(\nabla\phi\right)^2-2\left(\nabla\phi\right)^4\right)-\alpha_5
\mathrm{e}^{\phi}\left(\mathcal{G}-8G^{\mu\nu}\phi_\mu\phi_\nu-12\Box\phi
\left(\nabla\phi\right)^2-12\left(\nabla\phi\right)^4\right)\Biggr\},
\label{eq:complete}
\end{align}
where for short we have defined $\phi_\mu\equiv \partial_\mu\phi$.
One can see that the resulting action for
$\lambda_5=\beta_5=\alpha_5=0$ coincides with the theory
(\ref{eq:pot_fern}) presented in \cite{Fernandes:2021dsb}. We recall
that for this theory, the scalar field equation is conformally
invariant in 4 dimensions although the $\alpha_4$-dependent part of
the action {\it{is not}}. As regards the parts of the action
depending on $\lambda_5$, $\beta_5$, $\alpha_5$, they correspond to
the most general densities which are {\it{conformally invariant in
five dimensions}}, see for
example~\cite{Oliva:2011np,Giribet:2014bva}. This motivates our use
of the subscripts $4$ and $5$ for the parametrization of the full
action under consideration~(\ref{eq:complete}). The distant relation
of our full action to conformal invariance is rather surprising, as
at no point in our construction did we advocate conformal symmetry
of any sort.

As announced, the potentials (\ref{eq:pot_mu_egal_1}), along with
the scalar field (\ref{eq:choice_of_phi}), solve the compatibility
conditions (\ref{eq:comp1})--(\ref{eq:comp3}) for $\mu=1$, for two
distinct sets of relations holding between the coupling constants
and the constant appearing in the scalar field, thus yielding two
distinct metric solutions of the form (\ref{metric}). The first
solution exists with all coupling constants switched on, namely,
\begin{equation}
\lambda_4 = \frac{\beta_4^2}{4\alpha_4},\quad \lambda_5 =
\frac{9\beta_5^2}{20\alpha_5},\quad\frac{\beta_5}{\beta_4} =
\frac{2\alpha_5}{3\alpha_4}.\label{eq:coupling_non_vanishing}
\end{equation}
The solution reads,
\begin{align}
\phi={}&\ln\left(\frac{\eta}{r}\right),\quad \eta = \sqrt{\frac{-2\alpha_4}{\beta_4}},\nonumber\\
f\left(r\right) = {} & 1 + \frac{2\alpha_5\eta}{3\alpha_4r}
+\frac{r^2}{2\alpha_4}\left[1\pm
\sqrt{\left(1+\frac{4\alpha_5\eta}{3r^3}\right)^2+4\alpha_4\left(\frac{\Lambda}{3}+\frac{2
M}{r^3}+\frac{2\alpha_4}{r^4}+\frac{8\alpha_5\eta}{5r^5}\right)}\right],\label{eq:sol_1}
\end{align}
where $M$ is a free integration constant. A number of comments can
be made concerning this solution which is of the Boulware-Deser
type~\cite{Boulware:1985wk} typical of higher order metric theories,
admitting two branches. The ``plus'' branch is asymptotically of
de-Sitter or anti de-Sitter type much like \cite{Boulware:1985wk}.
In Lovelock theory this upper branch is perturbatively
unstable~\cite{Charmousis:2008ce} although no direct analogy can be
made with the case here. We will not consider this branch any
further as we are mostly interested in asymptotically flat
spacetimes. So for simplicity let us set $\Lambda=0$ and consider
the ``negative'' branch in order to discuss some properties of the
solution. For the negative branch we have a Schwarzschild  limit as
the coupling constants $\alpha_4$, $\alpha_5$ tend to zero. Also,
for $\alpha_5=0$, which automatically implies $\beta_5=\lambda_5=0$,
the first class of solutions of Ref. \cite{Fernandes:2021dsb} is
recovered. The asymptotics $r\to\infty$ of the full solution are
given by,
\begin{equation}
f(r)=1-\frac{2M}{r}-\frac{2\alpha_4}{r^2}-\frac{8\alpha_5\eta}{5r^3}+\mathcal{O}\left(\frac{1}{r^4}\right),
\end{equation}
and $M$ is therefore the ADM mass. The function $f$ has the same
behavior as the Schwarzschild metric up to first order while the
next order is controlled by the coupling $\alpha_4$, and the other
couplings $\alpha_5$ and $\beta_4$ appear via $\eta$ in the higher
corrections. When we have a horizon the black hole has therefore
secondary hair, as the only integration constant appearing in the
metric is $M$, while all other constants are fixed by the theory.
Note that even if $M=0$, the spacetime is not trivial, and is in
fact a black hole or naked singularity. On the other hand, as $r\to
0$, the metric function (\ref{eq:sol_1}) behaves as
\begin{equation}
f(r)=\left\lbrace\begin{aligned}& -\frac{1}{5}-\frac{21\alpha_4 r}{50\alpha_5\eta}+\mathcal{O}\left(r^2\right)\quad\text{if }\alpha_5>0\\ & \frac{4\alpha_5\eta}{3\alpha_4 r}+\frac{11}{5}+\mathcal{O}\left(r\right)\quad\text{if }\alpha_5<0\end{aligned}\right. ,
\end{equation}
and while $f(0)$ is finite for $\alpha_5>0$, spacetime curvature is
infinite at $r=0$ since $f(r)$ does not possess a regular de Sitter
core $f(r)=1+\delta r^2+o\left(r^2\right)$ (with $\delta$ a
constant). In fact, the spacetime might not be defined in the whole
$r\in\left(0,\infty\right)$ but only on $\left(r_S,\infty\right)$
where $r_S>0$ is such that the square root ceases to be well-defined
below $r_S$. This branch singularity is quite typical for Lovelock
spacetimes \cite{Charmousis:2008kc}. Before discussing the horizon
structure of the solution~(\ref{eq:sol_1}), let us present the
second $\mu=1$ solution arising from action (\ref{eq:complete}).

Indeed a second, quite distinctive class of solutions exists,
provided the couplings of the generalized conformal action of
\cite{Fernandes:2021dsb} are switched off,
\begin{equation}
\lambda_4=\beta_4=\alpha_4=0,\quad
\lambda_5=\frac{9\beta_5^2}{20\alpha_5},\label{eq:coupling_vanishing}
\end{equation}
with a scalar field and a metric function given by
\begin{equation}
\phi=\ln\left(\frac{\eta}{r}\right),\quad \eta =
2\sqrt{\frac{-\alpha_5}{3\beta_5}},\quad
f\left(r\right)=\frac{1}{1+\frac{4\alpha_5\eta}{3r^3}}\left[1-\frac{\Lambda
r^2}{3}-\frac{2M}{r}-\frac{4\alpha_5\eta}{15
r^3}\right],\label{eq:sol_2}
\end{equation}
where $M$ is a mass integration constant and $\Lambda$ the
cosmological constant which we again set to zero for simplicity.
Note that although the action includes higher order terms, these do
not yield branching solutions as is common in Lovelock theories
\cite{Boulware:1985wk}. In fact the solution is very much "GR like".
Indeed, asymptotically as $r\to\infty$, the metric function behaves
as
\begin{equation}
f(r) = 1-\frac{2M}{r}-\frac{8\alpha_5\eta}{5r^3}+\mathcal{O}\left(\frac{1}{r^4}\right).
\end{equation}
On the other hand, close to the origin the metric function does not
blow up and behaves as
\begin{equation}
f(r) =-\frac{1}{5}-\frac{3Mr^2}{2\alpha_5\eta}+\mathcal{O}\left(r^3\right).
\end{equation}
Note that although the $r^2$ term is what is needed for a regular
black hole (see for example \cite{Hayward:2005gi}), regularity is
spoiled by $f(0)=-1/5$. Hence at $r=0$ we have a curvature
singularity. If $\alpha_5<0$, the spacetime becomes singular at
$0<r_S = \left(-4\alpha_5\eta/3\right)^{1/3}$, unless the numerator
of (\ref{eq:sol_2}) also vanishes at $r_S$, which occurs for a mass
$M=M_S$ with
\begin{equation}
M_S =\frac{6^{2/3}\left(-\alpha_5\eta\right)^{1/3}}{5}.\label{eq:ms}
\end{equation}
If $\alpha_5>0$ on the other hand, we always have a black hole horizon even for the case $M=0$. In fact one can see that for $\alpha_5>0$ and constant $M$ the horizon position is shifted at larger $r$, in comparison to GR  i.e., $r_h>2M$. One can also show that there is an unstable light ring at $r_L>3M$ again at larger $r$ in comparison to GR (see for example \cite{EventHorizonTelescope:2020qrl}).

\begin{figure}[!htb]
\begin{subfigure}{6cm}
\includegraphics[width=\linewidth]{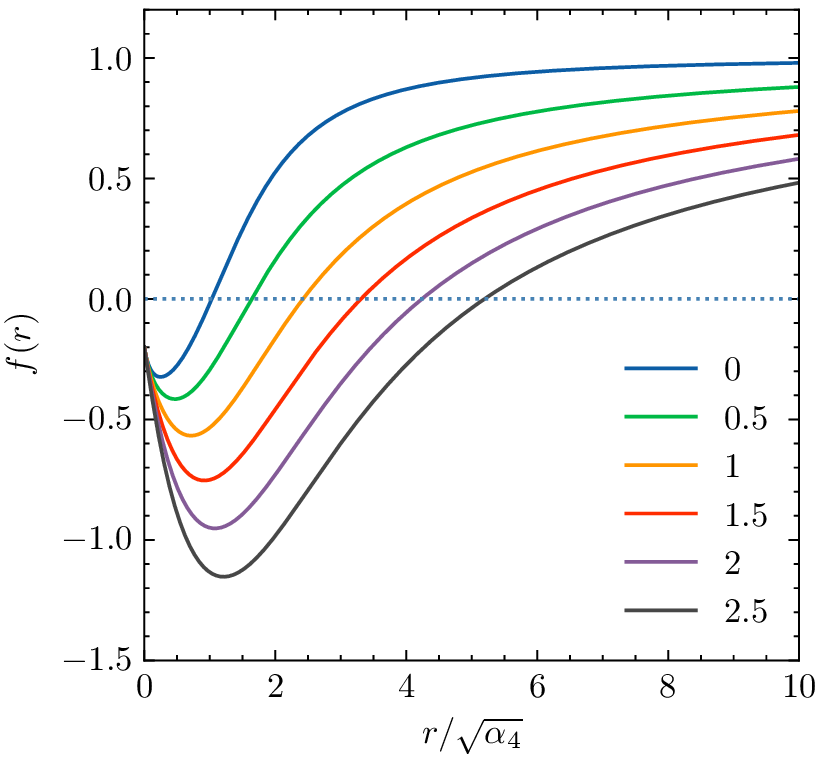}
\end{subfigure}
\begin{subfigure}{5.8cm}
\includegraphics[width=\linewidth]{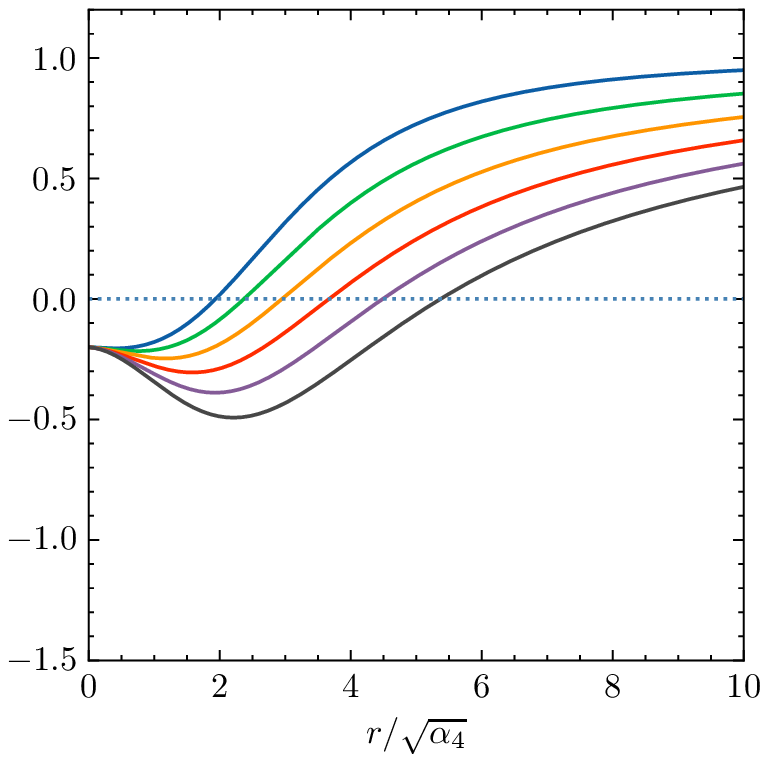}
\end{subfigure}
\begin{subfigure}{5.8cm}
\includegraphics[width=\linewidth]{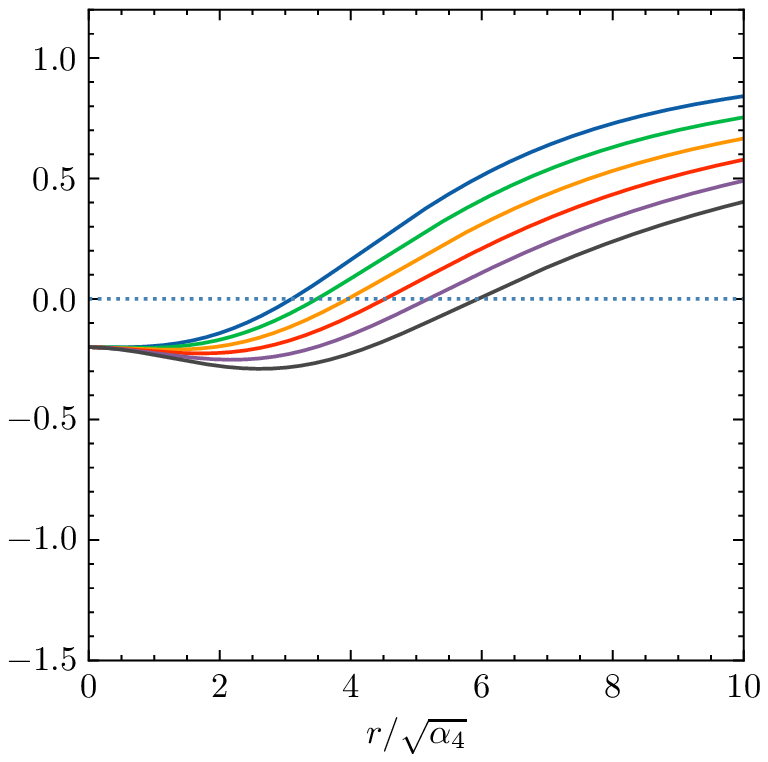}
\end{subfigure}
\caption{Metric profile $f(r)$ of Eq. (\ref{eq:sol_1}) for
$\alpha_4>0$ and different values of the mass (in units of
$\sqrt{\alpha_4}$, indicated by the colors) and different positive
values of the product $\alpha_5\eta$ (in units of $\alpha_4^{3/2}$),
namely: on the left $\alpha_5\eta=0.25$, in the middle
$\alpha_5\eta=20$, on the right $\alpha_5\eta=100$. The spacetime is
a black hole for any mass, with hidden singularity at $r=0$.}
\label{fig:1}
\end{figure}

\begin{figure}[!htb]
\begin{subfigure}{6.1cm}
\includegraphics[width=\linewidth]{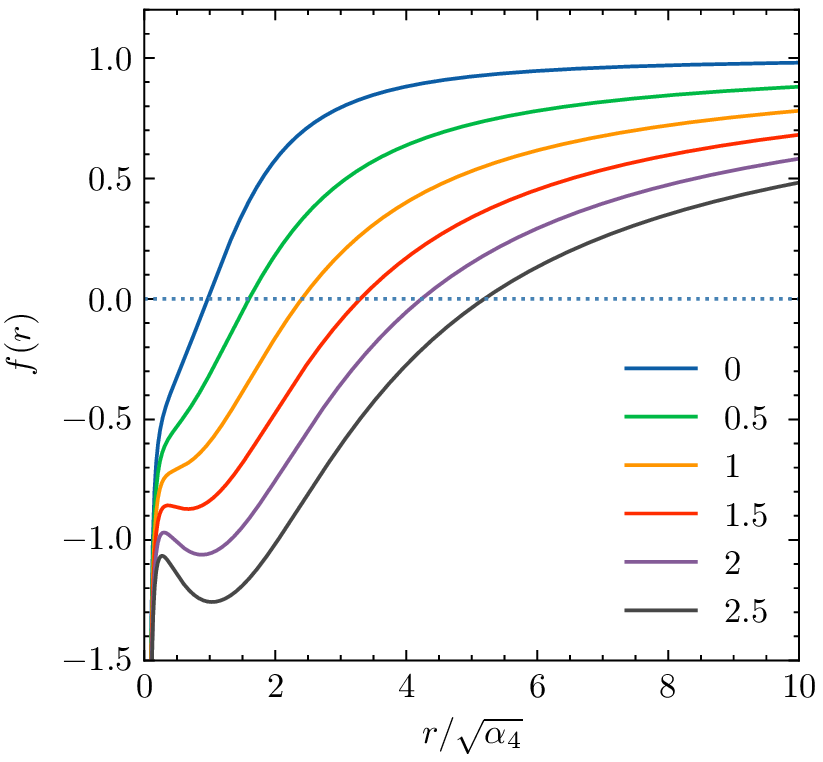}
\end{subfigure}
\begin{subfigure}{5.8cm}
\includegraphics[width=\linewidth]{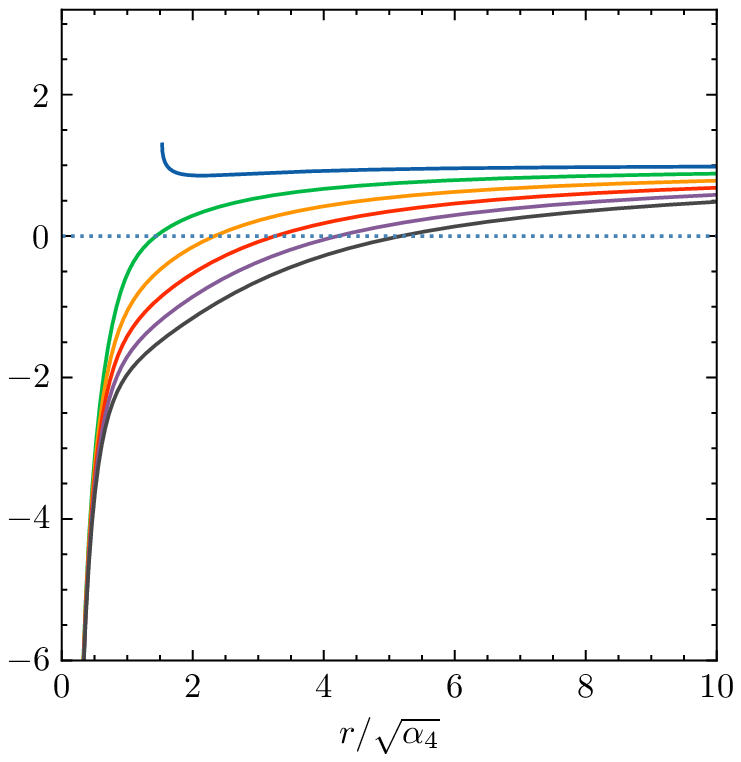}
\end{subfigure}
\begin{subfigure}{5.8cm}
\includegraphics[width=\linewidth]{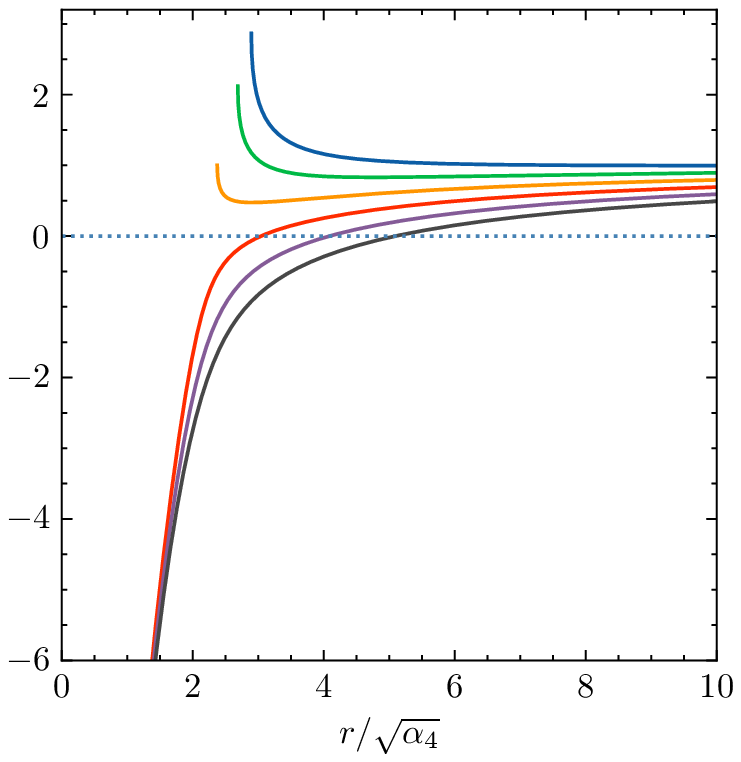}
\end{subfigure}
\caption{Metric profile $f(r)$ of Eq. (\ref{eq:sol_1}) for
$\alpha_4>0$ and different values of the mass (in units of
$\sqrt{\alpha_4}$, indicated by the colors) and different negative
values of the product $\alpha_5\eta$ (in units of $\alpha_4^{3/2}$),
namely: on the left $\alpha_5\eta=-0.25$, in the middle
$\alpha_5\eta=-2$, on the right $\alpha_5\eta=-10$. On the left, the
spacetime is a black hole for any mass. When
$\left\lvert\alpha_5\eta\right\rvert$ increases (middle and right
plots), the light spacetimes acquire a naked singularity at a radius
$r_S>0$, while the heavier spacetimes remain black holes with hidden
singularity at $r=0$.} \label{fig:2}
\end{figure}

\begin{figure}[!htb]
\begin{subfigure}{6.1cm}
\includegraphics[width=\linewidth]{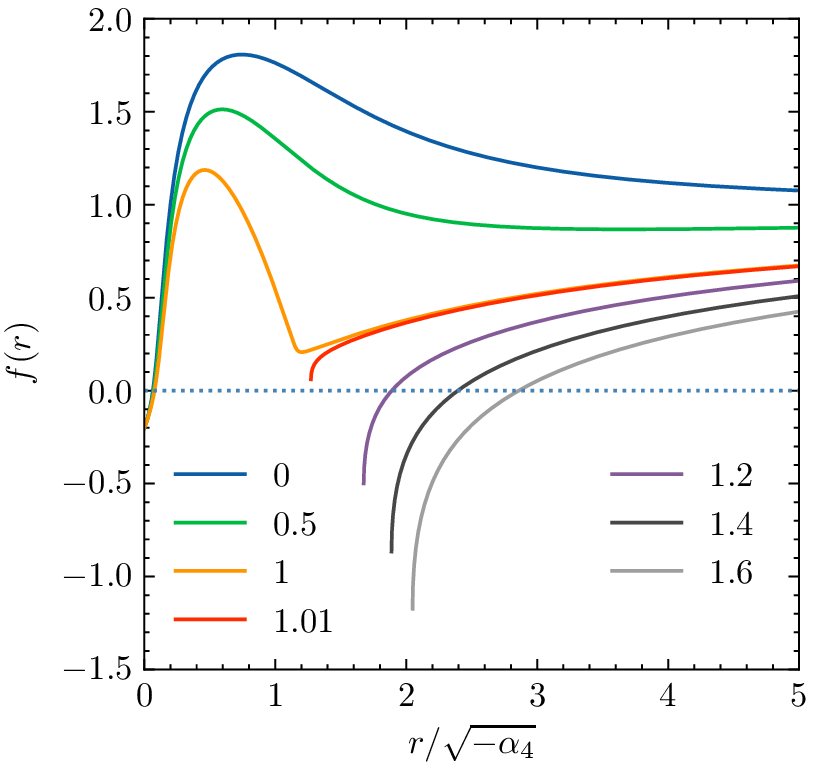}
\end{subfigure}
\begin{subfigure}{5.8cm}
\includegraphics[width=\linewidth]{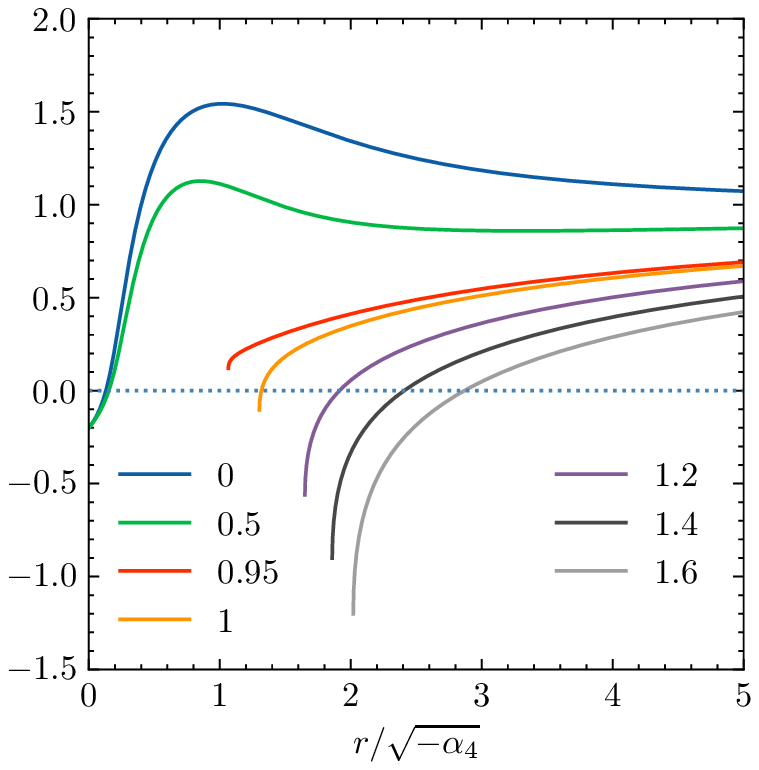}
\end{subfigure}
\begin{subfigure}{5.8cm}
\includegraphics[width=\linewidth]{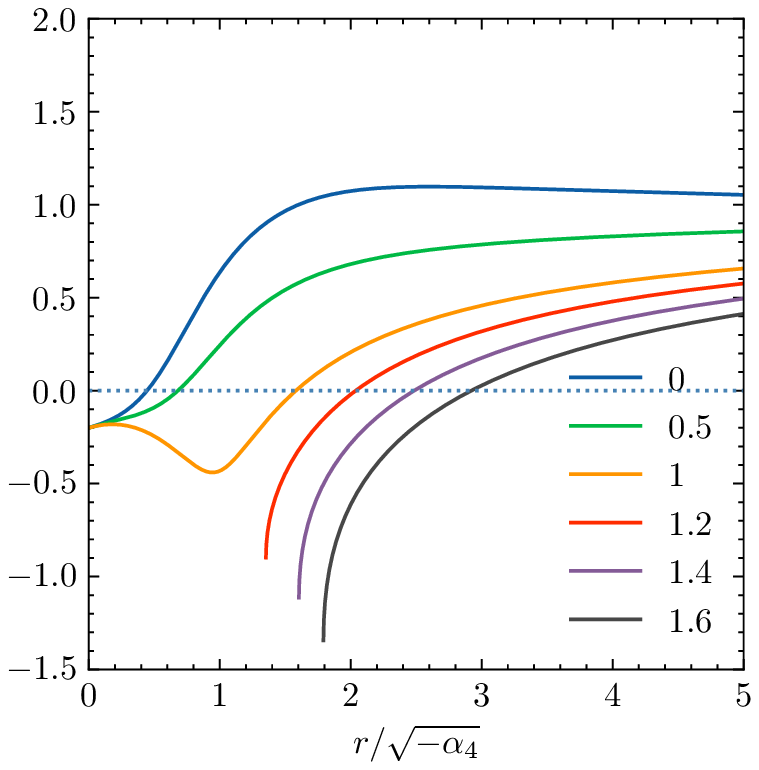}
\end{subfigure}
\caption{Metric profile $f(r)$ of Eq. (\ref{eq:sol_1}) for
$\alpha_4<0$ and different values of the mass (in units of
$\sqrt{-\alpha_4}$, indicated by the colors) and different positive
values of the product $\alpha_5\eta$ (in units of
$\left(-\alpha_4\right)^{3/2}$), namely: on the left
$\alpha_5\eta=0.25$, in the middle $\alpha_5\eta=0.5$, on the right
$\alpha_5\eta=2$. For small $\alpha_5\eta$ (left and middle plots),
light and heavy masses give black holes, with hidden singularity at
$r=0$ or at $r_S>0$ respectively, while intermediate masses give a
naked singularity at $r_S>0$ (see the red curves on the left and
middle plots). For sufficiently large $\alpha_5\eta$ (right plot),
all spacetimes are black holes, with hidden singularity at $r=0$ for
light masses or at $r_S>0$ for large masses.} \label{fig:3}
\end{figure}

\begin{figure}[!htb]
\begin{subfigure}{6.1cm}
\includegraphics[width=\linewidth]{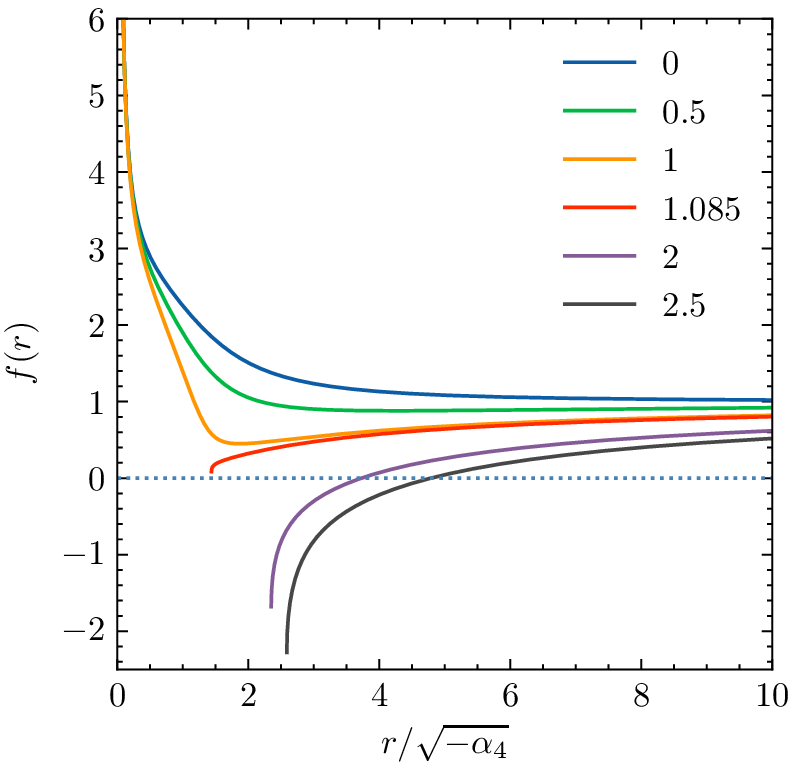}
\end{subfigure}
\begin{subfigure}{5.8cm}
\includegraphics[width=\linewidth]{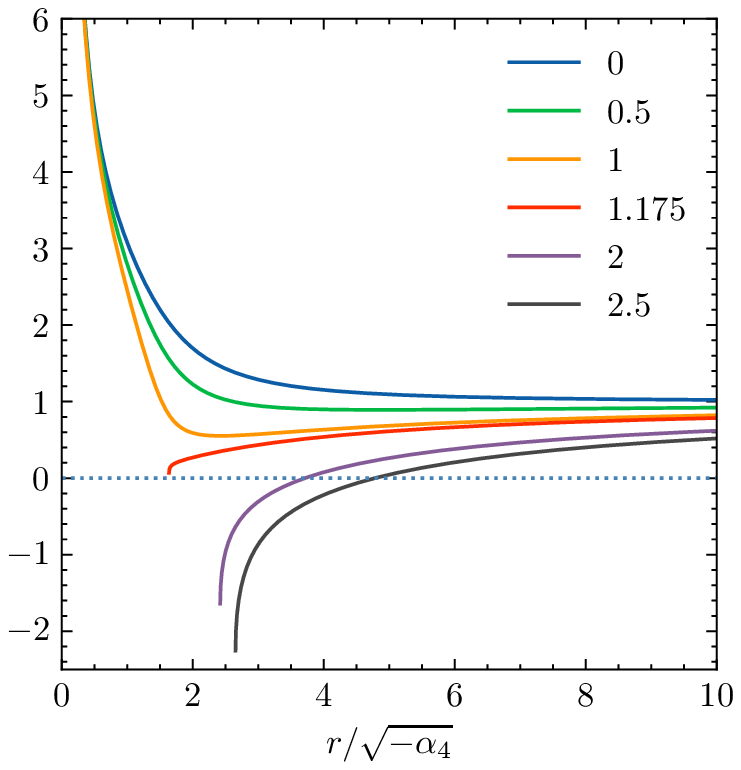}
\end{subfigure}
\begin{subfigure}{5.8cm}
\includegraphics[width=\linewidth]{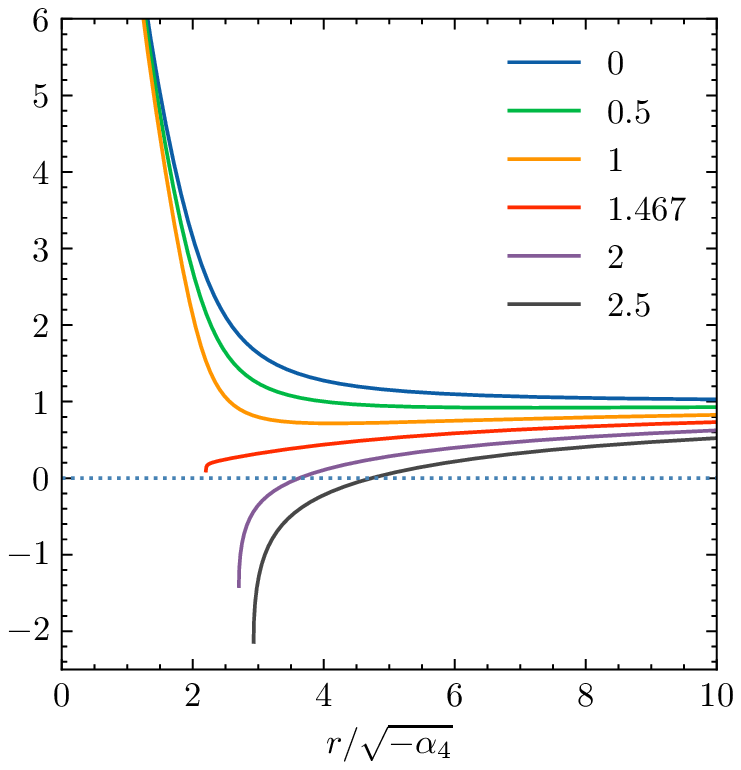}
\end{subfigure}
\caption{Metric profile $f(r)$ of Eq. (\ref{eq:sol_1}) for
$\alpha_4<0$ and different values of the mass (in units of
$\sqrt{-\alpha_4}$, indicated by the colors) and different negative
values of the product $\alpha_5\eta$ (in units of
$\left(-\alpha_4\right)^{3/2}$), namely: on the left
$\alpha_5\eta=-0.25$, in the middle $\alpha_5\eta=-1$, on the right
$\alpha_5\eta=-5$. The spacetime is a naked singularity at $r=0$ for
light masses, a naked singularity at $r_S>0$ for intermediate masses
(see the red curves), and a black hole with hidden singularity at
$r_S>0$ for large masses.} \label{fig:4}
\end{figure}

\begin{figure}[!htb]
\begin{subfigure}{6.3cm}
\includegraphics[width=\linewidth]{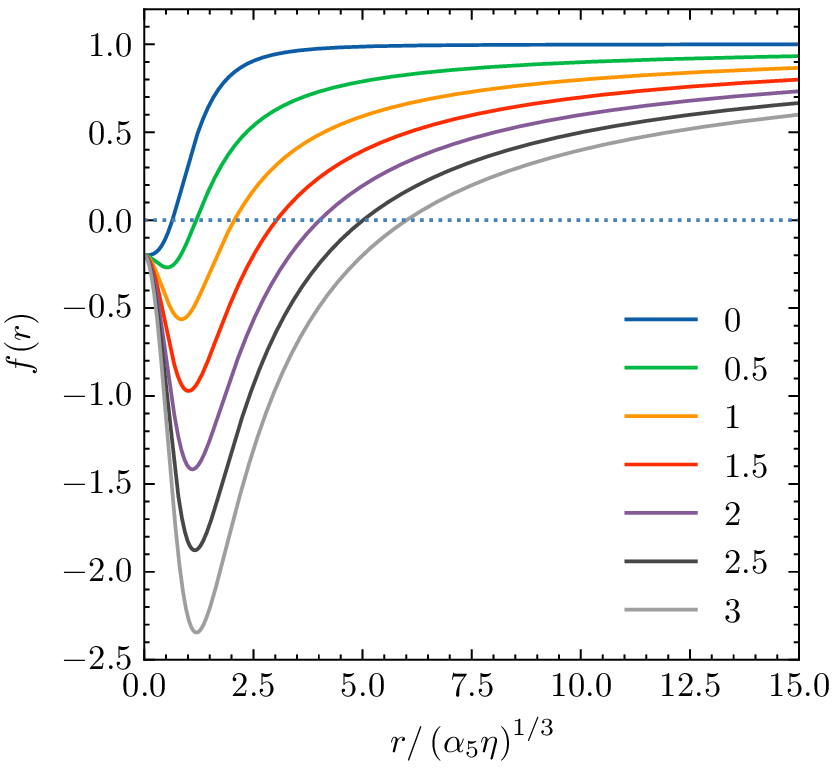}
\end{subfigure}
\begin{subfigure}{5.8cm}
\includegraphics[width=\linewidth]{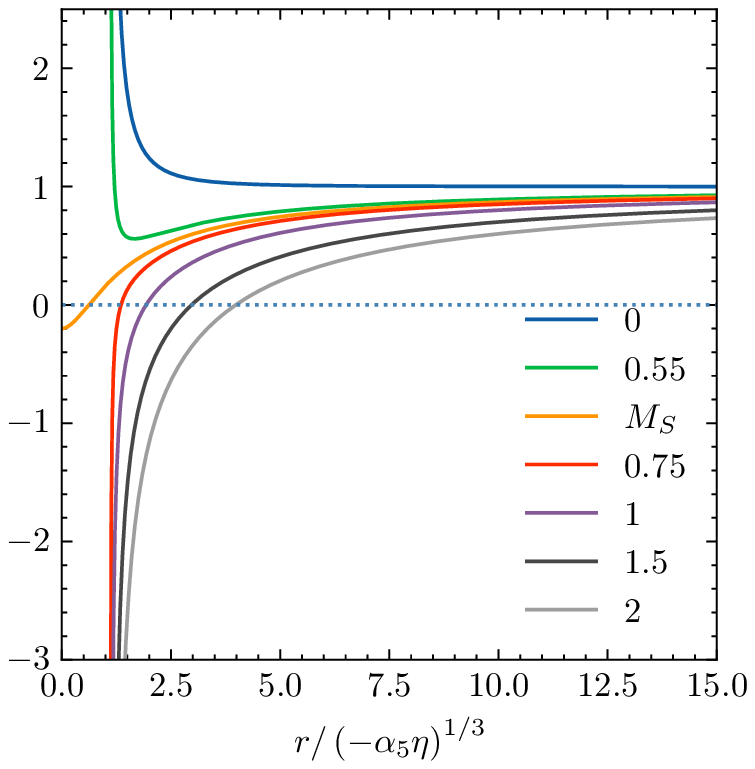}
\end{subfigure}
\caption{Metric profile $f(r)$ of Eq. (\ref{eq:sol_2}) for
$\alpha_5>0$ (left plot) or $\alpha_5<0$ (right plot) and different
values of the mass (in units of
$\left\lvert\alpha_5\eta\right\rvert^{1/3}$, indicated by the
colors). For $\alpha_5>0$, the spacetime is a black hole for any
mass. For $\alpha_5<0$, the only mass giving a singularity at $r=0$
is $M_S$, see Eq. (\ref{eq:ms}), and this singularity is hidden by a
horizon. For other masses, there is a singularity at
$r_S=\left(-4\alpha_5\eta/3\right)^{1/3}$, either naked for $M\leq
M_S$ or hidden by a horizon if $M\geq M_S$.} \label{fig:5}
\end{figure}

More generally, as regards the horizon of both spacetimes (\ref{eq:sol_1}) and (\ref{eq:sol_2}), they turn out to be given by a cubic polynomial equation,
\begin{equation}
15r_h^3-30Mr_h^2-15\alpha_4r_h-4\alpha_5\eta=0,\label{eq:equation_horizon}
\end{equation}
where $r=r_h$ is the location of the event horizon, and with of course $\alpha_4=0$ in the case of (\ref{eq:sol_2}). This condition is necessary, but not sufficient. It is sufficient in the case of (\ref{eq:sol_1}) if $\alpha_4>0$ and $\alpha_5>0$, and in the case of (\ref{eq:sol_2}) if $\alpha_5>0$, or if $\alpha_5<0$ and $M> M_S$. In order to sketch the general aspect of the spacetimes and their horizons, we present various plots of the functions $f(r)$ of Eqs.~(\ref{eq:sol_1}) and (\ref{eq:sol_2}). Regarding solution~(\ref{eq:sol_1}), note that the subcase $\alpha_5=\beta_5=\lambda_5=0$ was studied in~\cite{Babichev:2022awg} where it was found that there is a bias towards negative values of $\alpha_4${\footnote{This is based on the observation of atomic nuclei while assuming a Birkhoff type argument for this theory. It is also based on the observation that the canonical kinetic term has the usual sign for $\alpha_4<0$.}}. The plots in
Figs.~\ref{fig:1}, \ref{fig:2}, \ref{fig:3} and \ref{fig:4} present respectively the cases ($\alpha_4>0$, $\alpha_5>0$),
($\alpha_4>0$, $\alpha_5<0$), ($\alpha_4<0$, $\alpha_5>0$) and
($\alpha_4<0$, $\alpha_5<0$). The obtained spacetimes have at most
one horizon. It is only when $\alpha_4$ and $\alpha_5$ are positive that there is always a horizon (even for $M=0$). This is due to the fact that the square root is never zero and no branch singularity is possible. For all the
other cases however we may have naked singularities for certain values
of the coupling constants. An exotic result, in the case of Fig.
\ref{fig:3}, left and middle plots, is a mass gap and horizon gap
between light black holes and heavy black holes: masses
$M\in\left(M_1,M_2\right)$ give rise to naked singularities while
$M< M_1$ or $M> M_2$ give black holes, the black holes with
$M< M_1$ having very tiny horizons. It is interesting to note, in comparison with \cite{Babichev:2022awg}, that for $\alpha_4<0$ we may do away with cases of naked singularities by having sufficiently positive $\alpha_5$ (right plot of Fig.~\ref{fig:3}). Concerning~(\ref{eq:sol_2}),
its profile is presented in Fig.~\ref{fig:5}: if $\alpha_5>0$, it is
a black hole for any mass, while for $\alpha_5<0$, it is a black
hole only for $M\geq M_S$ where $M_S$ is given by (\ref{eq:ms}).

The two cases where a horizon exists for any mass, that is to say, Fig.~\ref{fig:1}, i.e. solution (\ref{eq:sol_1}) with $\alpha_4>0$, $\alpha_5>0$, and Fig.~\ref{fig:5}, left plot, i.e. solution (\ref{eq:sol_2}) with $\alpha_5>0$, are in fact strongly constrained by the following argument, which was formerly developed in \cite{Charmousis:2021npl} and assumes that the considered solutions verify a Birkhoff type argument, more precisely, that they are the unique static, spherically-symmetric solutions of their respective theories. In this case, these solutions must represent the gravitational field created by an atomic nucleus, of radius $R\sim 10^{-15}\,$m and mass $M\sim 10^{-54}\,$m. Since these nuclei can be experimentally probed, they are not covered by a horizon, and therefore $r_h<R$ where $r_h$ is root of Eq. (\ref{eq:equation_horizon}). It is easy to show that, for the considered two cases, this leads to the constraints
\begin{equation}
0<\alpha_4<R\left(R-2M\right)\sim 10^{-30}\,\text{m}^2,\quad 0<\alpha_5\eta<\frac{15}{4}R^2\left(R-2M\right)\sim 10^{-45}\,\text{m}^3,
\end{equation}
where of course the first inequality does not concern Fig. 5, left plot, which already has $\alpha_4=0$. Such stringent bounds make the associated gravitational effects probably undetectable.

In the case of solution (\ref{eq:sol_2}), thus remains most likely a unique case, $\alpha_5<0$ (Fig.~\ref{fig:5}, right plot), where black holes have a minimal mass $M_S$ given by (\ref{eq:ms}). Another argument from \cite{Charmousis:2021npl} can then be used to constrain the value of $\left\lvert\alpha_5\eta\right\rvert$ for this theory. Indeed, the minimal mass $M_S$ must be lower than the mass of experimentally detected black holes. In GW200115, the second object is a black hole of mass $M=5.7^{+1.8}_{-2.1}M_\odot$ at $90\%$ credible interval, giving,
\begin{equation}
\left\lvert\alpha_5\eta\right\rvert \lesssim 2070^{+565}_{-659}\,\text{km}^3.
\end{equation}
If we take into account other events for which the second object is lighter but whose black hole nature is not certain, namely GW170817 and GW190814, we rather get $\left\lvert\alpha_5\eta\right\rvert \lesssim 230\,\text{km}^3$ and $\left\lvert\alpha_5\eta\right\rvert \lesssim 194\,\text{km}^3$. Finding such simple constraint would not be possible in the other cases, of Figs.~\ref{fig:2}, \ref{fig:3} and \ref{fig:4}, since in these cases, the minimal mass depends non trivially on both $\alpha_4$ and $\alpha_5\eta$.

This completes our study of solutions of the form (\ref{metric}) to
action (\ref{eq:complete}) with a nontrivial scalar field that has a
typical logarithmic running. From a string theoretical point of view
this is to be corrected in the UV from higher order corrections, but
one can also question the existence of solutions with a constant
scalar field $\phi=\phi_0$. A more general analysis of such
solutions is displayed in Appendix A. It is easy to see that, for
the considered action (\ref{eq:complete}), a constant scalar field
solution exists provided the couplings satisfy
\begin{align}
0={}&\alpha_4^2\left[\alpha_4^3 \alpha_5 (\beta_4 \lambda_5-2 \beta_5 \lambda_4)+
\alpha_4^4 \beta_5 \lambda_5-5 \alpha_4 \alpha_5^3 \lambda_5+4 \alpha_5^4 \lambda_4\right]+2\alpha_5^5\left(2\alpha_5\beta_4-3\alpha_4\beta_5\right)\Lambda , \label{eq:cond_cst}\\
0\neq{}&\alpha_5^3-\alpha_4^2\alpha_5\beta_4+\alpha_4^3\beta_5,\label{eq:cond_cst_2}
\end{align}
and the solution is a Schwarzschild-(A)dS black hole,
\begin{equation}
\phi_0 = \ln\left(-\frac{\alpha_4}{\alpha_5}\right),\quad
f(r)=1-\frac{2M}{r}+\frac{\alpha_4^4\left(\alpha_4\lambda_5-\alpha_5\lambda_4\right)-
\Lambda\alpha_5^5}{3\alpha_5^2\left(\alpha_5^3-\alpha_4^2\alpha_5\beta_4+\alpha_4^3\beta_5\right)}r^2.\label{eq:sads}
\end{equation}
Interestingly enough, this solution is valid for the
theory with coupling constants given by
(\ref{eq:coupling_non_vanishing}) where we also have the black hole solution (\ref{eq:sol_1}). However, while (\ref{eq:sol_1}) is asymptotically flat if $\Lambda=0$, this is not the case of (\ref{eq:sads}) which
has an effective cosmological constant.

\section{Case of constant $\mu\not=1$}
In the previous section we have analyzed in some detail the action
and solutions for $\mu=1$  obtained from integrating
(\ref{eq:comp1})--(\ref{eq:comp3}). Using the method of Sec. II we
now proceed to solve the potentials of our theory $V$, $W$, $Z$ from
(\ref{eq:comp1})--(\ref{eq:comp3}) for $\mu$ an arbitrary constant.
Without loss of generality we have set $d=-1$ and $\tilde{c}=0$ and by
taking into account the scalar field
profile~(\ref{eq:choice_of_phi}), we can find the potentials as
functions of $\phi$,
\begin{align}
V ={}& \frac{\alpha}{c^{2(1-\mu)}}
(\mathrm{e}^{2\left(1-\mu\right)\phi} -1)- \frac{\gamma}{c}
(2\mu+1)(\mu-1) \mathrm{e}^\phi,\nonumber\\
W+1 ={}& 2(\mu-1)c^{2(\mu-1)}
\mathrm{e}^{2\left(1-\mu\right)\phi}
\left(1+\frac{2\alpha}{c^2}\mathrm{e}^{2\phi}+
\frac{2\gamma }{c^{2\mu+1}}\mathrm{e}^{(2\mu+1)\phi}\right),\label{eq:pot}\\
Z={}&2(\mu-1)c^{2(\mu-1)}\mathrm{e}^{2\left(1-\mu\right)\phi}\left(
\frac{-2\Lambda\mu(2\mu+1)}{3(2\mu-1)}+
\frac{2(\mu-1)}{c^2}\mathrm{e}^{2\phi}
+\frac{2\alpha(\mu-2)}{c^4}\mathrm{e}^{4\phi}-\frac{12\gamma}{(2\mu+3)c^{2\mu+3}}
\mathrm{e}^{(2\mu+3)\phi}\right).\nonumber
\end{align}
Putting aside non-relevant constants, the potentials end up depending on four parameters $\alpha$, $\gamma$, $c$ and $\Lambda$ or rather coupling constants of the theory.
In appendix C, the cases $\mu=\pm \frac{1}{2}$ and
$\mu=\pm\frac{3}{2}$ are treated separately since they
correspond to some degenerate characteristic equations with metric
solutions involving a logarithmic radial dependence\footnote{A convenient overall rescaling of the potentials $V$, $W+1$ and $Z$ by $\tilde{V}=\frac{V}{2(\mu-1)c^{2(\mu-1)}
\mathrm{e}^{2\left(1-\mu\right)\phi}}$, etc.,  gives back the $\mu\rightarrow 1$ potentials of the previous section
(\ref{eq:pot_mu_egal_1}) with couplings related by (\ref{eq:coupling_non_vanishing}), up to redefinitions of the coupling constants.}.

Before proceeding to the discussion of the solutions, let us write down the resulting action and provide an interesting geometrical interpretation. To this end we introduce the following three Lagrangian densities,
\begin{eqnarray}
\label{lagconf} {\cal L}_1(n)&=&2 e^{n\phi},\qquad {\cal L}_2(n)=
e^{(n-2)\phi}\left[R+(n-1)(n-2)\left(\nabla\phi\right)^2\right],\nonumber\\
{\cal L}_3(n)&=&
e^{(n-4)\phi}\left[\mathcal{G}-4(n-3)(n-4)G^{\mu\nu}\nabla_\mu\phi\nabla_\nu\phi-2(n-2)(n-3)(n-4)
\Box\phi\left(\nabla\phi\right)^2 \right.\\
&&\left. -(n-2)(n-3)^2(n-4)\left(\nabla\phi\right)^4\right].\nonumber
\end{eqnarray}
The actions corresponding  to the first two Lagrangians ${\cal L}_1$ and ${\cal L}_2$ are conformally invariant in dimension $n>2$,
while for ${\cal L}_3$ the conformal invariance holds for dimension $n>4$. The resulting action for $\mu\not=1$ can be conveniently written as a linear combination of
the densities (\ref{lagconf}) as\footnote{For simplicity, we have
set the cosmological constant $\Lambda=0$, and normalized the full
action by a global factor for latter convenience.}
\begin{eqnarray}
\label{actionmu} &&S_{\mu\not=1}=\frac{1}{2}\int
d^4x\sqrt{-g}\Bigg\{\alpha\left[2(\mu-2)c^{2(\mu-3)}\,{\cal
L}_1(6-2\mu)+4 c^{2(\mu-2)}\,{\cal
L}_2(6-2\mu)+\frac{c^{2(\mu-1)}}{(\mu-1)}\,{\cal
L}_3(6-2\mu)\right]\nonumber\\
&&-\gamma\left[\frac{12}{(2\mu+3)\,c^5}\,{\cal
L}_1(5)-\frac{4}{c^3}\,{\cal L}_2(5)+\frac{(2\mu+1)}{c}\,{\cal
L}_3(5)\right]+2(\mu-1) c^{2(\mu-2)}\,{\cal L}_1(4-2\mu)+2
c^{2(\mu-1)}\,{\cal L}_2(4-2\mu) \Bigg\}.
\end{eqnarray}
As one can see, the ``integrable'' four dimensional action we have
obtained for constant $\mu\not=1$ turns out to be a linear
combination of densities that {\it{are conformally invariant}} in
dimensions $D=5$, $D=6-2\mu$ and $D=4-2\mu$.
Strictly speaking, and given the current restriction $\mu\not\in\left\lbrace\pm 1/2,\pm 3/2\right\rbrace$,
this interpretation holds true if $\mu=-\frac{k}{2}$ with $k\in \mathbb{N}\setminus\left\lbrace 1,3\right\rbrace$. Case $\mu=1$ can also be recovered from (\ref{actionmu}) in a
non-trivial way. Indeed, all terms of (\ref{actionmu}) but one have
a well-defined limit as $\mu\to 1$, and notably, the last term gives
a contribution $\mathcal{L}_2\left(4-2\mu\right)\to
\mathcal{L}_2\left(2\right)$, yielding the standard
Einstein-Hilbert piece. As regards the \textit{a priori} singular term $\frac{c^{2(\mu-1)}}{(\mu-1)}\,{\cal L}_3(6-2\mu)$,
a singular limiting procedure as described in Ref.~\cite{Babichev:2023rhn} yields precisely the $\alpha_4$--part of action (\ref{eq:complete}). Another particular value is $\mu=0$, which corresponds in (\ref{actionmu}) to densities which are conformally invariant in
dimensions $D=4$, $D=5$ and $D=6$. However, even with proportionality factor null, $\mu=0$, in
(\ref{eq:comp1})--(\ref{eq:comp3}) the below presented metric
solutions, (\ref{eq:f_generic}) and (\ref{eq:linear_case}), can be
verified to be solutions to the corresponding $\mu=0$ theory, and
therefore, all comments made on these solutions are also valid for
this degenerate case.

Note that for the special case $\gamma=0$, the theory, and its
solution presented below, see~(\ref{eq:f_generic}), correspond to
the Galileon theory and black hole found
in~\cite{Charmousis:2012dw}. This Galileon black hole has a higher
dimensional (i.e. non-Newtonian) fall off, as it originates from a
higher dimensional Lovelock solution with a horizon given as a
product of two-spheres. More precisely, it comes from a diagonal
Kaluza-Klein reduction along an internal space which is a product of
$s=\mu -1$ two-spheres. This higher-dimensional interpretation holds
strictly speaking for any integer $\mu\geq 2$, while the other
values of $\mu$ are just an analytic continuation of this result. In
particular, the identified favorable case, $\mu = 1$, corresponds
formally to an empty product of two-spheres, which explains that it
can be captured by the Kaluza-Klein reduction only through a
singular limit. In a nutshell, in action (\ref{actionmu}), the parts
which have no $\gamma$ factor have two possible, complementary
interpretations: in terms of Lagrangians which have conformal
invariance in dimensions $6-2\mu$ and $4-2\mu$, or in terms of
Kaluza-Klein reduction. Interestingly, the former interpretation
seems more relevant when $\mu\leq 0$, while the latter
interpretation makes more sense for $\mu\geq 1$. As regards the
$\gamma$ parts, they are just the Lagrangians with conformal
invariance in five dimensions, and their interpretation is the same
for any $\mu$.

Having identified the geometric nature of the action, we now
concentrate on the solutions of the theory~(\ref{actionmu}).
The quadratic equation satisfied by $f$ can be written generically
for any $\mu\notin \left\lbrace \pm3/2,\pm 1/2\right\rbrace$ as
\begin{equation}
\alpha f^2-\frac{f}{2\mu-1}\left(r^2+2 \gamma
r^{1-2\mu}+2\alpha\right)+\frac{1}{(2\mu-1)^2} \left(r^2
-\frac{\Lambda r^4}{3}-2M r^{3-2\mu}-\frac{2 \gamma
(2\mu-1)}{2\mu+3}r^{1-2\mu}+\frac{\alpha(2\mu-1)}{2\mu-3}\right)=0,
\label{eq:poly}
\end{equation}
and its solution is given by
\begin{equation}
f(r)=\frac{1}{2\mu-1}\left[ 1+ \frac{\gamma
r^{1-2\mu}}{\alpha}+\frac{r^2}{2\alpha}\left( 1\pm
\sqrt{\left(1+\frac{2\gamma}{
r^{2\mu+1}}\right)^2+\frac{4\alpha\Lambda}{3}+\frac{8\alpha
M}{r^{2\mu+1}}+\frac{16 \alpha \gamma
(2\mu+1)}{(2\mu+3)r^{2\mu+3}}-\frac{8
\alpha^2}{(2\mu-3)r^4}}\right)\right].\label{eq:f_generic}
\end{equation}
It is easy to see that in the limiting case $\mu=1$, the expression
of $f$ reduces to the $\mu=1$ solution previously found in~(\ref{eq:sol_1}) with $\alpha=\alpha_4$ and $\gamma=2\alpha_5\eta/3$. For brevity, we discuss these solutions when we switch
off the potential term $\Lambda$ (which played the role of the
cosmological constant for the $\mu=1$ solution). The asymptotic
behaviour at $r\to\infty$ of the metric function depends on the
range of the parameter $\mu$, and also on the sign of $\gamma$
($>0$, $<0$ or $=0$) with respect to the sign in front of the square
root. If $\mu>-1/2$, the metric potential $f$ grows at infinity with
a power exponent that never exceeds~2. Indeed in this case  for the
lower branch we have asymptotically either $f\sim(2\mu-1)^{-1}$ (for
$\mu>1/2$) or $f\sim r^{1-2\mu}$ (for $\mu<1/2$), while for the
upper branch $f\sim r^2$. On the other hand, for $\mu<-1/2$, there
are three cases. If $\gamma$ has a sign opposite to the sign in
front of the square root in~(\ref{eq:f_generic}), then $f\sim r^2$.
In the case when those signs coincide (or when $\gamma =0$), $f$ has
an asymptotic dependence on $r$ with the power which exceeds that of
(A)dS, namely $f\sim r^{1-2\mu}$ (or $f\sim r^{3/2-\mu}$).

This is a fairly general case of a black hole solution where the
solution has a unique integration constant, the mass $M$, and
two coupling constants $\alpha$ and $\gamma$ appearing in the
metric. It is a black hole with secondary hair. We can note that the
metric solution does not depend on the coupling $c$ while the
parameter $\Lambda$ loosely parameterizes the Liouville potential
which is the higher dimensional cosmological constant for
non-critical string theories \cite{Dixon:1986iz}. The theory can be
perceived as a higher order Liouville theory  with higher order
corrections parameterized here by $\alpha$ and $\gamma$. It is also
clear that depending on the ranges of the constant $\mu\not=1$, the
metric function will display different asymptotic behaviors, and no
one of them will reproduce the standard fall off $f\sim 1-2M/r$. The
"mass" contribution as $r\to\infty$ is for instance $\sim
Mr^{1-2\mu}$ for $\mu>-1/2$. Because of that, one can wonder
whether these solutions have a finite mass. In fact, a quick
thermodynamic analysis of the solution through the Euclidean method
on the action (\ref{actionmu}) reveals that the mass ${\cal M}$ is
finite for both branches, and parameterized in terms of the
proportionality factor $\mu$ as
$$
{\cal M} = \frac{\mu}{2\mu-1}M,
$$
or in term of the horizon $r_h$,
$$
{\cal M}=
\frac{\mu}{2}\left[\frac{\alpha}{(2\mu-3)}r_h^{2\mu-3}+\frac{1}{(2\mu-1)}r_h^{2\mu-1}-\frac{2\gamma}{(2\mu+3)r_h^2}\right].
$$

Similar to the $\mu=1$ case, for $\alpha=0$, the quadratic
equation~(\ref{eq:poly}) reduces to a linear equation for $f$ whose solution is
given by
\begin{equation}
f(r)=\frac{1}{(2\mu-1)\left(1+\frac{2\gamma}{r^{2\mu+1}}\right)}\left[1-\frac{\Lambda
r^2}{3}-\frac{2M}{r^{2\mu-1}}-\frac{2\gamma(2\mu-1)}{(2\mu+3)r^{2\mu+1}}\right].\label{eq:linear_case}
\end{equation}
The corresponding theory is given by the linear combination of the
conformal densities of the dimensions $D=5$ and $D=4-2\mu$, and for
$\mu=1$, one recovers the previous solution (\ref{eq:sol_2}), along
with its theory (\ref{eq:complete}) with couplings
(\ref{eq:coupling_vanishing}). Note that the solution (\ref{eq:linear_case}) and its associated action both remain valid for $\gamma=0$. Also, as in the
quadratic case, the asymptotic behavior of the metric solution
(\ref{eq:linear_case}) depends on the parameter $\mu$.

\section{Conclusions}

In this paper we have considered quite general Horndeski theories,
without any apparent symmetries, which admit interesting and
explicit black hole solutions. Our starting point was a subclass of
Horndeski theory with arbitrary $\phi$-dependent potentials and
higher order operators; the latter are related to Kaluza-Klein
reduction of Lovelock theory (see for example
\cite{Charmousis:2014mia} and references within). We then, for
spherical symmetry, established integrability
(\ref{eq:simple_equation}) and compatibility relations
(\ref{eq:comp1})--(\ref{eq:comp3}) that permitted the selection of
interesting and quite general subsets of Horndeski theories without
apparent symmetries. There is no particular bias in the Horndeski
functionals for the filtered out theory; for example there is no
assumption of shift or parity symmetry and in four dimensions, no
conformal invariance of the scalar field equation. The compatibility conditions we found, by fully integrating the
field equations, involved an arbitrary function $\mu(r)$. We
established the repertory of some previously known solutions in
terms of our potentials and this function $\mu$. The BBMB~\cite{BBM} or MTZ~\cite{Martinez:2002ru}
solutions are found to yield quite complex functions $\mu(r)$
whereas those of Fernandes \cite{Fernandes:2021dsb} are tailored to
$\mu=1$. We then chose $\mu$ to be constant, for simplicity, in the
principal part of our study. Our filtered theories are at the end
parameterized in terms of this constant parameter and four coupling
constants. The solutions we have found are rare examples of unbiased Horndeski
functionals along with a canonical kinetic term for the scalar. They
have a higher dimensional geometric interpretation either through
conformally coupled actions (of various dimensions) or Kaluza-Klein
reduction of Lovelock theories. The physical properties of the
solutions depend on the value of $\mu$ which essentially fixes the
asymptotic behavior.

In the case $\mu=1$ we have a four dimensional Newtonian fall-off,
and the theory and solutions present the most phenomenological
interest. In this case the overall theory has distinct subparts with
four and five dimensional (conformal)
symmetries associated to the coupling of the scalar field while no symmetry is associated to the full action in four dimensions. The first of these sub actions, $\mathcal{L}_4(g,\phi)$, is a singular limit giving a
conformally coupled scalar field equation in four dimensions
\cite{Fernandes:2021dsb}. The second is a Lagrangian density ${\cal
L}_5(g,\phi)$ which is conformally invariant in $D=5$ dimensions.
Summing up, the full action in the case $\mu=1$ is given by the
Einstein-Hilbert plus the linear combination of the aforementioned
Lagrangians, ${\cal L}_4(g,\phi)$ and ${\cal L}_5(g,\phi)$. The
solutions emanating from this action are asymptotically flat (or
(A)dS in the presence of a cosmological constant in the action). The
black holes are asymptotically very similar to Schwarzschild,
depending on a mass parameter and two coupling constants associated
to the aforementioned 'four' and 'five-dimensional' parts in the
action. Each of these two parts of the action accompanied by the
Einstein-Hilbert term, possesses black hole solutions, and, somewhat
surprisingly, their combination also admits a black hole which, in a
sense, is a superposition of the latter two. Usually, because of
non-linearity, solutions of different actions cannot be superposed,
nevertheless the black holes we presented in this case can be viewed
as a non linear superposition of sub-solutions. If we switch off the
${\cal L}_4(g,\phi)$ part of the action we find a very much like
Schwarzschild type solution although the action itself is of higher
order. The vacua of such theories are not trivial and we have made
additional analysis in this direction, see appendix~D.

In fact for all remaining constant $\mu$ the action we consider
involves again two subparts:  the five dimensional piece ${\cal
L}_5(g,\phi)$ is a common denominator for all $\mu$. The second
piece can be identified in two ways:  as a Kaluza-Klein reduction
originating from a higher dimensional Lovelock theory or
interestingly, as a combination of conformally coupled actions in
differing dimensions. The black holes for the general action will
have, not surprisingly given their higher order nature, a
non-Newtonian fall off for the mass of the solution. The solution is
again a superposition of two differing solutions for the two subset
actions.  A relevant question is: does there exist an oxidation of
the full solution we have found into vacuum Lovelock metric theory
or are these solutions only scalar-tensor? Furthermore, could it be
that the five dimensional piece is some form of higher dimensional
matter \cite{Bardoux:2010sq} for a Lovelock black hole~\cite{Bogdanos:2009pc}? This is the most probable outcome as in
Lovelock theory there are strong constraints on the possible horizon
metrics which are permissible \cite{Dotti:2005rc}.

The solutions we have found in the $\mu=1$ case bifurcate known no hair
theorems, either due to the absence of symmetry in the action or due to the
presence of higher order corrections. Let us comment on a no hair theorem that concerns a two derivative scalar-tensor action of the form
\begin{eqnarray}
S=\int d^4x\,\sqrt{-g}\left[\kappa R-\beta {\cal
L}_2(n)-U(\phi)\right], \label{nohairaction}
\end{eqnarray}
where ${\cal L}_2(n)$ defined in (\ref{lagconf}) corresponds to the
non minimally coupled scalar. Here $\kappa$ and $\beta$ are
constants and $U$ is a function that only depends on the
scalar field. As mentioned before, in four dimensions, the density
${\cal L}_2(n)$ is conformally invariant only for $n=4$, and in this
case, a black hole solution for $U\equiv 0$ is known \cite{BBM}. Generalizing this solution for a nonminimal coupling parameter
$n\not=4$ remains an open problem. Regarding this question, a
no-hair theorem was established in Ref. \cite{Mayo:1996mv}
stipulating the non-existence of asymptotically flat black hole
solutions with a positive semidefinite potential for
$n\in[0,2[-\left\{1\right\}$. In the present case, our solution
(\ref{eq:sol_2}) arises from an action of the form
(\ref{nohairaction}) with $n=5$ and with $U\propto e^{5\phi}$ but
supplemented by the Lagrangian ${\cal L}_3(5)$ as defined in
(\ref{lagconf}). Our theory violates the hypotheses of the no-hair
theorem due to the presence of the higher order terms in ${\cal
L}_3(5)$. Nevertheless, the value of $n$, for our solution $n=5$, is
not forbidden by the no-hair theorem. Therefore one may wonder if
one could extend the theorem of~\cite{Mayo:1996mv} to include higher
order corrections.

There are several ways to extend our present analysis. For a start
we chose to take $\mu=\text{cst.}$ to facilitate the integration of
compatibility equations (\ref{eq:comp1})--(\ref{eq:comp3}). Other
choices however are possible with the BBMB and MTZ solutions
recovered for non constant $\mu$ as we mentioned earlier. As these
examples show, it is actually intriguing that once $\mu$ is not
constant then the function invariably depends on the mass parameter
of the solution. This renders the solution quite different; for
example it is known that the BBMB solution explodes at the horizon
of the black hole while for mass equal to zero the scalar is
trivial. Furthermore, although the mass parameter enters in the
coupling constants of the scalar field, these are in turn
accommodated in the theory, so that the mass never appears in the
action. Note that in the contrary this would have meant a fine
tuning of the theory for a given mass black hole!  Are these facts
accidents or more general observations? Furthermore, our
integrability condition was adapted to a homogenous spherically
symmetric ansatz for which $-g_{tt}=g^{rr}$. In fact it is not so
clear how to extend the integrability condition for a general
spherically symmetric ansatz. Is it possible to go beyond this and
establish again more general integrability conditions?

The integrability condition we have found admits a second branch of possible solutions which we did not manage to solve for more general theories than those found previously
\cite{Fernandes:2021dsb}, see appendix B. In
this second branch the scalar field has explicit dependence on the
metric function as well as it incorporates the flat vacuum.
Furthermore it incorporates an additional constant independent of
the mass which one may associate to the symmetry of the conformally
coupled scalar in \cite{Fernandes:2021dsb}. It is then normal to
question if the absence of extensions of the second branch is due to
the absence of conformal symmetry for the scalar: the scalar field
equation arising from the variation of the action (\ref{actionmu})
is conformally invariant only for $\mu=1$ and $\gamma=0$. This
intuition is comforted by our results in higher dimension where we
have shown that this second branch indeed exists for conformally
coupled scalar field~\cite{Babichev:2023rhn}. Another interesting question concerns the possible presence of multiple solutions and the issue of scalarisation \cite{Doneva:2017bvd}, for a {\it{given specific}} theory. Our unbiased Horndeski theories possibly allow for scalarisation and obtaining explicit, rather than numerical solutions, is an intriguing possibility. This is an interesting subject as well as undertaking issues of cosmological stability in such scalarised cases \cite{Anson:2019uto}. This and some of the
points mentioned above deserve further study which we hope will be
successfully undertaken in the near future.

\acknowledgments

We would like to thank Eloy Ay\'on-Beato and Karim Noui for
interesting discussions. We are greatful to ANR project COSQUA for
partially supporting the visit of CC in Talca Chile, where this work
was initiated. The work of MH has been partially supported by
FONDECYT grant 1210889. EB and NL acknowledge support of ANR grant
StronG (ANR-22-CE31-0015-01). The work of NL is supported by the
doctoral program Contrat Doctoral Sp\'ecifique Normalien \'Ecole
Normale Sup\'erieure de Lyon (CDSN ENS Lyon).

\appendix

\section{Constant scalar field}
In this appendix, we solve the field equations obtained from action
(\ref{eq:action}) for a constant scalar field
$\phi=\phi_0=\text{cst.}$ and a static, spherically-symmetric metric
ansatz as given in (\ref{metric}). For such a constant scalar field,
two cases must be distinguished, either $1+W\left(\phi_0\right)$
vanishes or not. Given the form of action (\ref{eq:action}), the
vanishing of $1+W\left(\phi_0\right)$ makes the Einstein-Hilbert
term disappear, which is equivalent to having all potentials (but
$W$) in the action (\ref{eq:action}) displaying infinite coupling
constants. In other words, $1+W\left(\phi_0\right)=0$ amounts to a
strongly coupled constant scalar field.

Let us start by the most relevant case, where the scalar field is
not strongly coupled, $W\left(\phi_0\right)\neq -1$. Then, the
metric field equations $\mathcal{E}_{\mu\nu} = \frac{\delta
S}{\delta g^{\mu\nu}}$ are solved by a Schwarzschild-(A)dS
metric profile, with the cosmological constant
determined by the constant scalar field,
\begin{equation}
f(r) =
1-\frac{2M}{r}+\frac{Z\left(\phi_0\right)r^2}{6\left(1+W\left(\phi_0\right)\right)},\label{eq:f_phi_constant_1}
\end{equation}
where $M$ is a mass integration constant. Entering this into the
scalar field equation $\mathcal{E}_\phi=\frac{\delta S}{\delta
\phi}$ gives
\begin{equation}
\Biggl[Z_\phi +
\frac{2Z}{1+W}\left(\frac{ZV_\phi}{3\left(1+W\right)}-W_\phi\right)+\frac{48M^2V_\phi}{r^6}\Biggr]_{\phi=\phi_0}=0,\label{eq:dsdphi}
\end{equation}
and, hence this will be a solution  only if
$V_\phi\left(\phi_0\right)=0$ and
\begin{equation}
\Biggl[Z_\phi - \frac{2ZW_\phi}{1+W}\Biggr]_{\phi=\phi_0}=0.
\end{equation}

We now turn for completeness to the strongly coupled case,
$W\left(\phi_0\right)=-1$. Then the equations $\mathcal{E}_{\mu\nu}$
are degenerate and are solved if $Z\left(\phi_0\right)=0$, while
$\mathcal{E}_\phi$ then fully determines the metric, giving the
solution\footnote{If $V_\phi\left(\phi_0\right)=0$ and
$W_\phi\left(\phi_0 \right)= 0$, the field equations are degenerate
and reduce to $Z_\phi\left(\phi_0\right)=0$ without any constraint
on $f(r)$.},
\begin{equation}
f(r) = \left\lbrace\begin{aligned} &
1+\frac{r^2W_\phi}{4V_\phi}\left(1\pm\sqrt{1+\frac{8V_\phi}{W_\phi}\left(\frac{2M}{r^3}-\frac{q}{r^4}-\frac{Z_\phi}{12W_\phi}\right)}\right)
\quad\text{if } V_\phi\left(\phi_0\right)\neq 0\text{ and
}W_\phi\left(\phi_0\right)\neq 0 \\ & 1\pm\sqrt{1+2Mr-q-\frac{Z_\phi
r^4}{24V_\phi}} \quad\text{if } V_\phi\left(\phi_0\right)\neq
0\text{ and }W_\phi\left(\phi_0\right)= 0\\ &
1-\frac{2M}{r}+\frac{q}{r^2} +\frac{r^2 Z_\phi}{12W_\phi}\quad
\text{if } V_\phi\left(\phi_0\right)= 0 \text{ and
}W_\phi\left(\phi_0\right)\neq 0\end{aligned}\right. ,
\end{equation}
where it is implicit that the potentials and their derivatives are
evaluated at $\phi_0$. These profiles display two integration
constants, a mass $M$ and a kind of "charge" $q$.

We can for instance apply the previous analysis to the case of
\cite{Fernandes:2021dsb}, with the potentials given by Eq.
(\ref{eq:pot_fern}). Then $V_\phi=-\alpha\neq 0$, therefore
solutions are obtained only in the strongly coupled case
$W\left(\phi_0\right)=-1$, giving $\phi_0=-\ln(\beta)/2$, while the
condition $Z\left(\phi_0\right)=0$ gives $\lambda=-\Lambda\beta^2$,
and the metric is given by the first line in the above equation, in
agreement with the results presented in \cite{Fernandes:2021dsb}.

\section{Analysis of the second branch}
We study in this appendix the second branch identified in Sec. II by
requiring that the second factor of the factorization
(\ref{eq:simple_equation}) vanishes. Using (\ref{eq:vk}-\ref{eq:v3})
this amounts to re-write the equation as
\begin{equation}
r^2 W_\phi + 4 V_\phi -
4fV_\phi\left[1+2r\phi'\frac{V_{\phi\phi}-\frac{V_\phi}{d}}{V_\phi}+r^2\left(\phi'\right)^2\frac{\frac{2+d_\phi}{d^2}V_\phi+V_{\phi\phi\phi}-\frac{3}{d}V_{\phi\phi}}{2V_\phi}\right]=0.
\end{equation}
In this generic form, there is no possibility to solve this
equation, and an option would be to look for separability of this
equation, as it happens in the particular case of Ref.~\cite
{Fernandes:2021dsb}. For this purpose, let us consider $\mathcal{U}$
such that,
\begin{equation}
\mathcal{U}_\phi = \frac{V_{\phi\phi}}{V_\phi}-\frac{1}{d},\quad
\mathcal{U}_\phi^2 =
\frac{2+d_\phi}{2d^2}+\frac{V_{\phi\phi\phi}}{2V_\phi}-\frac{3V_{\phi\phi}}{2dV_{\phi}},
\end{equation}
then the latter equation takes the form,
\begin{equation}
f\left(1+r\mathcal{U}'\right)^2-\left(1+\frac{r^2W_\phi}{4V_\phi}\right)=0.
\end{equation}
By introducing $u = \frac{1}{\alpha} \exp\left(\mathcal{U}\right)$,
with $\alpha$ a coupling constant, one has $u' = u\,\mathcal{U}'$,
so that
\begin{equation}
f\left(Y'\right)^2 - \frac{Y^2}{r^2}\left(1+\frac{W_\phi}{4V_\phi
u^2}Y^2\right)=0,
\end{equation}
where $Y = ru$. Therefore, the equation is separable if the last
term in the parenthesis is a pure function of $Y$, that is to say,
there exists a coupling constant $\beta$ such that,
\begin{equation}
\frac{W_\phi}{4V_\phi u^2} = \frac{\beta}{2\alpha}.
\end{equation}
Compiling all these conditions, one gets the following compatibility
conditions
\begin{equation}
V_{\phi\phi\phi} =
2\frac{V_{\phi\phi}^2}{V_\phi}-\frac{d_{\phi}}{d^2}V_\phi-\frac{V_{\phi\phi}}{d},\quad
W_\phi=\frac{2\beta}{\alpha}\gamma^2V_\phi^3\exp\left(-2\int\frac{\mathrm{d}\phi}{d}\right),\quad
u=\gamma
V_\phi\exp\left(-\int\frac{\mathrm{d}\phi}{d}\right),\label{eq:comp_second}
\end{equation}
where $\gamma$ is an additional coupling constant. These conditions
enable us to solve the second branch for $Y=ru$, yielding the
following expression for $u$,
\begin{equation}
u =
\frac{\sqrt{-2\alpha/\beta}}{r\cosh\left(c\pm\int\frac{\mathrm{d}r}{r\sqrt{f}}\right)}\quad\text{if
}\beta\neq 0,\quad u = \exp\left(c+\int\frac{\pm
1-\sqrt{f}}{r\sqrt{f}}\mathrm{d}r\right)\quad\text{if }\beta= 0,
\end{equation}
where $c$ an integration constant. One can see that in the case of
\cite{Fernandes:2021dsb} with the coupling functions
(\ref{eq:pot_fern}), $u=\exp\left(\phi\right)$, and this reproduces
the solution found in this reference. More generally, for $d(\phi)=\text{cst.}$, the conditions (\ref{eq:comp_second}) can be integrated and, taking into account the remaining field equations, one gets the solution of \cite{Fernandes:2021dsb}, up to scalar field redefinition. Hence, one concludes that the
second parenthesis of (\ref{eq:simple_equation}) can be made a
separable equation if conditions (\ref{eq:comp_second}) are
verified, and these conditions are then easily tractable for
$d\left(\phi\right)=\text{cst.}$\footnote{We recall that
only for the first branch of solution, the function
$d\left(\phi\right)$ can be taken as a constant without any loss of
generality.}, in which case one retrieves solutions of
\cite{Fernandes:2021dsb} up to scalar field redefinitions. We have
not proven in all generality that this second branch cannot lead to
new metric solutions, either in the separable case by a judicious
choice of non-constant $d\left(\phi\right)$, or even in the
non-separable case, but after long study, it has not seemed us that
an analytic solution could be obtained in these complicated
configurations.

\section{Cases $\mu=\pm1/2$ and $\mu=\pm3/2$}
We complete the integration of compatibility conditions
(\ref{eq:comp1})--(\ref{eq:comp3}) for constant $\mu$ by turning in
this appendix to the special values $\mu\in \left\lbrace \pm3/2,\pm
1/2\right\rbrace$, not included in the analysis of Sec. IV. For such
values, some factors or powers vanish in the previously found
potentials (\ref{eq:pot}) and/or in the corresponding metric
solutions (\ref{eq:f_generic}) and (\ref{eq:linear_case}). This
gives rise to additional terms in $\phi$ in the potentials, and to
logarithmic terms $\ln\left(r/c\right)$ in the metric function. The
potentials can again be written in terms of four parameters
$\alpha$, $\gamma$, $c$ and $\Lambda$. For $\mu = 3/2$, the
potentials all coincide with (\ref{eq:pot}). For $\mu=-3/2$, $V$ and
$W$ also coincide, while,
\begin{equation}
Z =
-\frac{5\mathrm{e}^{5\phi}}{c^5}\left(\frac{\Lambda}{2}-\frac{5\mathrm{e}^{2\phi}}{c^2}-\frac{7\alpha\mathrm{e}^{4\phi}}{c^4}-13\gamma-12\gamma\phi\right).
\end{equation}
For $\mu=-1/2$, the potentials are given by,
\begin{equation}
V =
\frac{\alpha\mathrm{e}^{3\phi}}{c^3}-\frac{3\gamma\mathrm{e}^{\phi}}{2c},\quad
W+1 =
-\frac{3\mathrm{e}^{3\phi}}{c^3}\left(1+\frac{2\alpha}{c^2}\mathrm{e}^{2\phi}-2\gamma\phi\right),\quad
Z=-\frac{3\mathrm{e}^{3\phi}}{c^3}\left(-\frac{\Lambda}{6}-\frac{3+\gamma}{c^2}\mathrm{e}^{2\phi}-\frac{5\alpha\mathrm{e}^{4\phi}}{c^4}+\frac{6\gamma\phi\mathrm{e}^{2\phi}}{c^2}\right).
\end{equation}
For $\mu=1/2$, they read,
\begin{equation}
V =
\frac{2\alpha\phi\mathrm{e}^{\phi}}{c}+\frac{\gamma\mathrm{e}^{\phi}}{c},\quad
W+1 =
-\frac{\mathrm{e}^{\phi}}{c}\left(1+\frac{2\left(\alpha+\gamma\right)\mathrm{e}^{2\phi}}{c^2}+\frac{4\alpha\phi\mathrm{e}^{2\phi}}{c^2}\right),\quad
Z =
-\frac{\mathrm{e}^{\phi}}{c}\left(\frac{\Lambda}{3}-\frac{\mathrm{e}^{2\phi}}{c^2}-\frac{\left(6\gamma+7\alpha\right)\mathrm{e}^{4\phi}}{2c^4}-\frac{6\alpha\phi\mathrm{e}^{4\phi}}{c^4}\right).
\end{equation}
Once again, $f$ is given by a polynomial equation of degree $2$ if
$\alpha\neq 0$ and degree $1$ if $\alpha=0$. When possible, we keep
the following expressions as close to the generic solution
(\ref{eq:f_generic}) as possible and thus do not always replace
explicitly $\mu$ by its value. First, if $\alpha\neq 0$, $\mu=3/2$
gives,
\begin{equation}
f(r)=\frac{1}{2\mu-1}\left[ 1+ \frac{\gamma
r^{1-2\mu}}{\alpha}+\frac{r^2}{2\alpha}\left( 1\pm
\sqrt{\left(1+\frac{2\gamma}{
r^{2\mu+1}}\right)^2+\frac{4\alpha\Lambda}{3}+\frac{8\alpha
M}{r^{2\mu+1}}+\frac{16 \alpha \gamma
(2\mu+1)}{(2\mu+3)r^{2\mu+3}}-\frac{8
\alpha^2\ln\left(r/c\right)}{r^4}}\right)\right],
\end{equation}
$\mu=-3/2$ gives,
\begin{equation}
f(r)=\frac{1}{2\mu-1}\left[ 1+ \frac{\gamma
r^{1-2\mu}}{\alpha}+\frac{r^2}{2\alpha}\left( 1\pm
\sqrt{\left(1+\frac{2\gamma}{
r^{2\mu+1}}\right)^2+\frac{4\alpha\Lambda}{3}+\frac{8\alpha
M}{r^{2\mu+1}}+32 \alpha \gamma \ln\left(\frac{r}{c}\right)-\frac{8
\alpha^2}{(2\mu-3)r^4}}\right)\right],
\end{equation}
$\mu=-1/2$ gives,
\begin{equation}
f(r)=\frac{1}{2\mu-1}\left[ 1+ \frac{\gamma
r^2\ln\left(r/c\right)}{\alpha}+\frac{r^2}{2\alpha}\left( 1\pm
\sqrt{\left(1+2\gamma
\ln\left(\frac{r}{c}\right)\right)^2+\frac{4\alpha\Lambda\ln\left(r/c\right)}{3}+8\alpha
M-\frac{8 \alpha \gamma}{r^2}-\frac{8
\alpha^2}{(2\mu-3)r^4}}\right)\right],
\end{equation}
and $\mu=1/2$ gives,
\begin{align}
f(r) ={}& -\frac{1}{2}\Biggl[
1+2\left(\frac{\gamma}{2\alpha}-\ln\left(\frac{r}{c}\right)\right)+\frac{r^2}{2\alpha}\Biggl(
1\pm  \nonumber\\ {}&{}
\sqrt{\left(1+\frac{2\left(\gamma-2\alpha\ln\left(r/c\right)\right)}{r^2}\right)^2+\frac{4\alpha\Lambda}{3}+\frac{8\alpha
M}{r^2}+\frac{8\alpha\ln\left(r/c\right)}{r^2}+\frac{2\alpha\left(3\alpha+2\gamma\right)}{r^4}-\frac{8
\alpha^2\ln(r/c)}{r^4}}\Biggr)\Biggr].
\end{align}
On the other hand, let us consider $\alpha=0$. Then, if $\mu = 3/2$,
$f$ coincides with the formula (\ref{eq:linear_case}) for generic
$\mu$. For $\mu=-3/2$,
\begin{equation}
f(r)=\frac{1}{(2\mu-1)\left(1+\frac{2\gamma}{r^{2\mu+1}}\right)}\left[1-\frac{\Lambda
r^2}{3}-\frac{2M}{r^{2\mu-1}}+2\gamma
r^2\left(1-4\ln\left(\frac{r}{c}\right)\right)\right],
\end{equation}
for $\mu=-1/2$,
\begin{equation}
f(r)=\frac{1}{(2\mu-1)\left(1+2\gamma\ln\left(\frac{r}{c}\right)\right)}\left[1+\left(2\gamma-\frac{\Lambda
r^2}{3}\right)\ln\left(\frac{r}{c}\right)-\frac{2M}{r^{2\mu-1}}+2\gamma
\right],
\end{equation}
and for $\mu=1/2$,
\begin{equation}
f(r)=-\frac{1}{2\left(1+\frac{2\gamma}{r^2}\right)}\left[1-\frac{\Lambda
r^2}{3}-2M+\frac{\gamma}{r^2}-2\ln\left(\frac{r}{c}\right)\right].
\end{equation}
The behaviour at $r\to\infty$ and $r\to 0$ is accordingly modified
with logarithmic terms as compared to generic $\mu$, but the
qualitative picture is the same as for the generic $\mu$ case of
Sec. IV: no Newtonian mass term is obtained.

\section{Stealth flat spacetime solutions to the action (\ref{eq:complete}) with a nontrivial scalar field $\phi=\phi(r)$}
In Sec. III, we have presented two new classes of black hole
solutions (\ref{eq:sol_1}) and (\ref{eq:sol_2}) provided the coupling
constants of action (\ref{eq:complete}) satisfy respectively (\ref{eq:coupling_non_vanishing}) or
(\ref{eq:coupling_vanishing}), and these solutions turn out to be
asymptotically flat for a vanishing bare cosmological constant
$\Lambda=0$. In this appendix, we would like to investigate whether
these asymptotically flat black holes coexist with some stealth flat
spacetimes. To this aim, we insert the ansatz (\ref{metric}) with
$f(r)=1$ in the field equations of action~(\ref{eq:complete}),
taking $\Lambda=0$. The combination of field equations
$\mathcal{E}^t_t-\mathcal{E}^r_r=0$ admits as well a similar
factorization to (\ref{eq:simple_equation}), which now reads,
\begin{equation}
\Bigl\{\left(\phi'\right)^2-\phi''\Bigr\}\Bigl\{r\mathrm{e}^{2\phi}
\left[2\beta_4+3\beta_5\mathrm{e}^{\phi}\right]-4\phi'
\left[\alpha_4\left(2+r\phi'\right)+\alpha_5\mathrm{e}^{\phi}\left(4+3r\phi'\right)\right]\Bigr\}=0.\label{eq:flat_eq}
\end{equation}
By opting for the second branch, one finds that there is a one
parameter family of theories, parameterized by a real number
$\epsilon$, such that a flat spacetime solution exists. The coupling
constants read
\begin{equation}
\lambda_4 =
\frac{3\epsilon\left(4-\epsilon\right)\beta_4^2}{16\alpha_4},\quad
\lambda_5=\frac{3\beta_5^2}{4\alpha_5},\quad
\beta_5=\frac{\epsilon\beta_4\alpha_5}{\alpha_4},
\end{equation}
while the  scalar field is given by
\begin{equation}
\phi =
\ln\left(\frac{16\alpha_4\alpha_5\left(2-\epsilon\right)}{r^2\alpha_4\beta_4\left(2-\epsilon\right)^2-16\epsilon\alpha_5^2}\right).
\end{equation}
Comparing the latter expression of $\lambda_5$ with its expression
required for the black holes solution,
(\ref{eq:coupling_non_vanishing}) or (\ref{eq:coupling_vanishing}),
one can see that both solutions cannot coexist. For the first
branch, one gets the same negative conclusion in which the stealth
solution is given by
\begin{equation}
\phi = \ln\left(\frac{\eta}{r}\right),\quad \eta =
\sqrt{\frac{4\alpha_5}{3\beta_5}},
\end{equation}
with the couplings satisfying
\begin{equation}
\beta_4 = 0,\quad \lambda_5 = \frac{9\beta_5^2}{4\alpha_5},\quad
\lambda_4=\frac{9\alpha_4\beta_5^2}{16\alpha_5^2}.
\end{equation}

\end{document}